\RequirePackage{fix-cm}

\documentclass[smallextended,natbib]{svjour3}       

\usepackage{amssymb}
\usepackage{amsmath}
\usepackage{framed}

\usepackage{amsfonts}
\usepackage{algpseudocode}
\usepackage{algorithm}
\usepackage{graphicx}
\usepackage{textcomp}

\usepackage{array}
\usepackage{rotating}
\usepackage{hyperref}
\usepackage{comment}
\usepackage{xspace}
\usepackage[table]{xcolor}
\usepackage{listings}
\usepackage{longtable}
\usepackage{blindtext}
\usepackage{lscape}
\usepackage{supertabular}
\usepackage{threeparttable}

\usepackage{pifont}
\usepackage{babel,booktabs,multirow}
\usepackage{lscape}
\usepackage{wasysym}
\usepackage{tikz}
\usepackage{longtable}
\usepackage{pdflscape}
\usepackage{fontawesome}
\usepackage{natbib}
\usepackage{marvosym} 

\usepackage[tight,footnotesize]{subfigure}

\usepackage{xspace}

\usepackage[colorinlistoftodos]{todonotes}

\newcommand*{\ie}{i.e.,\@\xspace}
\newcommand*{\eg}{e.g.,\@\xspace}

\newcommand*{\ME}{\texttt{Metagente}\@\xspace}
\newcommand*{\RM}{README.MD\@\xspace}

\newcommand\untick{\color{red}\ding{54}\color{black}}
\newcommand\tick{\color{green}\faCheck\color{black}}

\newcommand*{\tot}{17\@\xspace}

\newcommand*{\final}{10\@\xspace}

\newcommand\hand{\ding{45}}

\newcommand*{\GH}{GitHub\@\xspace}
\newcommand*{\CG}{ChatGPT\@\xspace}

\newcommand{\TST}{\textit{TS$_{10}$}\@\xspace}
\newcommand{\TSF}{\textit{TS$_{50}$}\@\xspace}
\newcommand{\ES}{\textit{ES}\@\xspace}

\lstdefinestyle{searchstringstyle}{
	basicstyle=\ttfamily\scriptsize,
	breaklines=true,                 
	captionpos=b,                    
	numbers=none,                    
	numbersep=5pt,                  
	showspaces=false,                
	showstringspaces=false,
	showtabs=false,                  
	tabsize=2,
	frame=single
}

\newcommand*{\CA}{\texttt{Prompt Creator Agent}\@\xspace}
\newcommand*{\EA}{\texttt{Extractor Agent}\@\xspace}
\newcommand*{\SA}{\texttt{Summarizer Agent}\@\xspace}
\newcommand*{\TA}{\texttt{Teacher Agent}\@\xspace}

\lstdefinestyle{pythonstyle}{
	language=Python,
	basicstyle=\footnotesize\ttfamily,
	keywordstyle=\color{blue}\bfseries,
	stringstyle=\color{red},
	commentstyle=\color{gray}\itshape,
	numbers=left,
	numberstyle=\tiny\color{gray},
	stepnumber=1,
	numbersep=5pt,
	backgroundcolor=\color{white},
	frame=single,
	rulecolor=\color{black},
	breaklines=true,
	breakatwhitespace=true,
	showstringspaces=false,
	columns=flexible,
	captionpos=b,
	xleftmargin=0.5em,
	xrightmargin=0.5em,
	aboveskip=1em,
	belowskip=1em,
	keepspaces=true,
	tabsize=4
}

\newcommand\revised[1]{\textcolor{blue}{#1}}

\newcommand\revOne[1]{\textcolor{blue}{#1}}
\newcommand\revTwo[1]{\textcolor{red}{#1}}
\newcommand\revThree[1]{\textcolor{orange}{#1}}

\newcommand\Comm[1]{\todo[size=\scriptsize, color=gray!60]{#1}}

\newcommand{\rqfirst}{\textbf{RQ$_1$}: \textit{To what extent do existing frameworks cover MAS foundational concepts?}}

\newcommand{\rqsecond}{\textbf{RQ$_2$}: \emph{What are the main characteristics of the considered MAS frameworks?}} 
\newcommand{\rqthird}{\textbf{RQ$_3$}: \textit{How well can popular frameworks support the summarization of \GH \RM files?}}

\definecolor{javagreen}{rgb}{0.25,0.5,0.35} 
\definecolor{javastring}{rgb}{0.6,0.125,0.02} 
\definecolor{javapurple}{rgb}{0.5,0,0.35} 
\definecolor{javadocblue}{rgb}{0.25,0.35,0.75} 

\lstdefinestyle{javaStyle}{
	language=Java,
	aboveskip=3mm,
	belowskip=3mm,
	fontadjust=true,
	frame = single,
	columns=flexible,
	basicstyle={\small\ttfamily},
	numbers=left,
	numberstyle=\tiny\color{gray},
	keywordstyle=\color{javapurple}\bfseries,
	commentstyle=\color{javagreen},
	stringstyle=\color{javastring},
	breaklines=true,
	breakatwhitespace=true,
	tabsize=3,
	showstringspaces=false,
	morekeywords={concept},
	morecomment=[s][\color{javadocblue}]{/**}{*/}
}

\newcommand{\mybox}[4]{
	\begin{figure}[h]
		\centering
		\begin{tikzpicture}
			\node[anchor=text,text width=\columnwidth-0.5cm, draw, rounded corners, line width=0.5pt, fill=#3, inner sep=1mm] (big) {\\#4};
			\node[draw, rounded corners, line width=.2pt, fill=#2, anchor=west, xshift=1mm] (small) at (big.north west) {#1};
		\end{tikzpicture}
	\end{figure}
}

\begin{document}

\sloppy

\title{Developing LLM-based Multi-Agent Systems in Software Engineering: A Mixed-Method Experience Report}


\titlerunning{Developing LLM-based MAS in SE: A Mixed-Method Experience Report}




\author{Mariama Celi Serafim De Oliveira \and
	Motunrayo Osatohanmen Ibiyo \and Marco Gianrusso \and
	Claudio Di Sipio \and
	Davide Di Ruscio \and
	Phuong T. Nguyen 
}


\institute{
Mariama Celi Serafim De Oliveira \at
DISIM, University of L'Aquila, 67100 L'Aquila, Italy \\
Åbo Akademi University, 20500 Turku, Finland \\
\email{mariamaceli.serafimdeoliveira@student.univaq.it}
\and Motunrayo Osatohanmen Ibiyo \at
DISIM, University of L'Aquila, 67100 L'Aquila, Italy \\
Åbo Akademi University, 20500 Turku, Finland \\
\email{motunrayoosatohanmen.ibiyo@student.univaq.it}
\and Marco Gianrusso \at
DISIM, University of L'Aquila, 67100 L'Aquila, Italy \\
\email{marco.gianrusso@univaq.it} 
\and Claudio Di Sipio \at
Johannes Kepler University, 4040 Linz, Austria \\
\email{claudio.di\_sipio@jku.at}
\and \Letter~Davide Di Ruscio \at
DISIM, University of L'Aquila, 67100 L'Aquila, Italy \\
\email{davide.diruscio@univaq.it}
\and Phuong T. Nguyen \at
DISIM, University of L'Aquila, 67100 L'Aquila, Italy \\
\email{phuong.nguyen@univaq.it}
}


\date{Received: date / Accepted: date}

\maketitle
\abstract{
	The proliferation of Generative Artificial Intelligence (Gen AI) powered by large language models (LLMs) has 
		transformed the software development process, introducing new paradigms for code generation, debugging, testing, and maintenance. While early applications focused on leveraging single, independent LLMs 
	to assist developers with isolated tasks, recent advances have shifted toward multi-agent systems (MAS) that orchestrate multiple LLM-based agents working collaboratively toward common objectives. 	
	Despite their promising potential, using MAS encompasses a set of challenges for developers who have to carefully select the right technology, devise proper coordination rules, and design specific roles for the involved agents. 	 
	In this paper, by means of a mixed-method study, we provide a comprehensive overview of the existing tools and frameworks for implementing MAS in software engineering. 
	First, we conducted a quantitative analysis of the most relevant open source MAS frameworks by evaluating their documentation, features, and capabilities from the developers' perspective. Second, we performed a qualitative evaluation of a subset of the selected frameworks by implementing a common use case, \ie the summarization of \GH \RM files. Our findings demonstrate that the selected frameworks provide a good coverage in terms of fundamental components of MAS, even though advanced features such as telemetry of agents are still missing. In addition, the empirical evaluation shows that there is no significant difference in terms of ROUGE scores considering the summarization task, while the time taken to complete the task varies significantly across the frameworks, suggesting that some frameworks are more suitable for rapid prototyping than others. Finally, we provide a set of lessons learned and challenges that can help researchers and practitioners to select a suitable MAS framework according to their needs.}










%

\section{Introduction}
\label{sec:Introduction}

The usage of large language models (LLMs) 
is pushing the boundaries of the automated software engineering, as different solutions are in place to support the whole development lifecycle, \eg from the requirement engineering \citep{10628487} to code development \citep{mastropaolo_studying_2021,BUCAIONI2024100526}. In this respect, these cutting-edge models contribute in defining a new software development paradigm, called AIware, in which LLMs are used as AI agents, being able to support developers in their daily tasks \citep{10.1145/3663529.3663820}. 


Although advanced LLMs outperform traditional approaches across many tasks, recent research reveals significant limitations that undermine the quality of their generated output. These limitations include: \emph{(i)} catastrophic forgetting~\citep{WANG2026132918}, in which where models lose previously learned knowledge when trained on new data; \emph{(ii)} hallucinations~\citep{LAVRINOVICS2025100844,10.1145/3703155,10.1145/3728894}, \ie generating plausible but factually incorrect or nonsensical content; \emph{(iii)} privacy concerns~\citep{KIBRIYA2024109698,BERINI2026104241}, \ie potential leakage of sensitive training data; and \emph{(iv)} limited reasoning capabilities for complex, multi-step problems. To overcome these limitations, LLM-based multi-agent systems (MAS) have emerged as a novel methodology to orchestrate, manage, and evaluate a set of AI agents for specific applications \citep{10.1145/3712003,li_survey_2024}. This transition represents a paradigm shift from monolithic LLM applications to the AgentWare paradigm, in which multiple LLMs are connected and work collaboratively to solve common tasks. By distributing responsibilities across specialized agents, MAS can leverage the strengths of individual models while mitigating their weaknesses through collaboration, verification, and iterative refinement. The adoption of LLM-based agents has accelerated rapidly in recent years~\citep{hou_large_2023}, with applications spanning diverse domains including software development, scientific research, and autonomous systems. However, despite the proliferation of MAS frameworks and implementations in Software Engineering (SE), a systematic and detailed comparison of their capabilities, performance, and suitability for specific SE tasks remains notably absent. This gap is particularly critical given the diversity of SE activities--ranging from code generation and debugging to requirements analysis and architectural design--each with distinct characteristics and quality requirements. Understanding how different MAS architectures perform across these varied tasks is essential for both researchers developing new agent systems and practitioners seeking to adopt these technologies effectively. 

In this paper, we adopted \emph{a mixed-method approach} \citep{10045_55590,7965402,10873003} with both qualitative and quantitative results to provide a comprehensive overview of the existing MAS frameworks, focusing on their capabilities and features. In particular, we first selected 20 existing frameworks from \GH and other sources. Then, we elicited the most popular and categorize them into two categories, \ie low-code and high-code frameworks, according to the level of abstraction they provide. We then analyzed the documentation and technical features of the selected frameworks, focusing on their capabilities to support the development of MAS. Finally, we conducted an empirical evaluation based on a common scenario, \ie the summarization of \GH README files, to assess the usability and effectiveness of the selected frameworks in practice. Our findings demonstrate that the elicited frameworks offers a good coverage in terms of fundamental components of MAS, \ie coordination rules, specification of roles and behaviors, and message handling. However, advanced features such as benchmarking, monitoring, and human-in-the-loop integration are still missing. On top of the obtained results, we delivered a set of lessons learned that can help researchers and practitioners to select the proper MAS framework according to their needs.

We aim to answer the following research questions (RQs). 


\begin{itemize}
\item \rqfirst~
To elicit the most advanced ones, we conducted a qualitative analysis using both the available documentation and previous studies. 

\item \rqsecond~Based on the identified frameworks, we performed an in-depth evaluation using a template-based approach. In particular, we focused on practical aspects that are relevant for developers, from the installation to the agent orchestration.

\item \rqthird~Once we selected the four most representative frameworks, we performed a quantitative analysis to evaluate their performance in a common use case, \ie the summarization of \GH \RM files.

\end{itemize}

In this respect, the contributions of this paper are summarized as follows:

\begin{itemize}
\item A comparative analysis of existing frameworks to develop LLM-based multi-agent systems.	
\item An empirical study to evaluate the effectiveness and efficiency of five frameworks for a common use case, \ie the summarization of \GH \RM files.
\item A set of lessons learned and practical guidelines for researchers and practitioners to develop MAS leveraging existing frameworks. 
\item A replication package including source code and the datasets formulated following existing practice~\citep{baltes2026guidelinesempiricalstudiessoftware} to foster future research in the domain \citep{replicationPackage}.
\end{itemize}


\noindent
\textbf{Structure.} Section~\ref{sec:RelatedWork} reviews the related work and background related to LLM-based 
MAS frameworks. 
Afterward, we present a qualitative analysis with selection of the frameworks and the evaluation process in Section~\ref{sec:Results}. Section \ref{sec:EmpiricalEvaluation} details an empirical study with the summarization of README files using MAS frameworks. A dedicated discussion concerning the lessons learned and practical guidelines for researchers and practitioners is provided in Section \ref{sec:Discussion}. 
Finally, we sketch future work and 
conclude the paper in Section~\ref{sec:Conclusion}.  

\section{Related Work}	
\label{sec:RelatedWork}

\subsection{LLMs-based Multi-Agent Systems in Software Engineering}

\begin{figure}[h]
	\centering
	\includegraphics[width=0.95\linewidth]{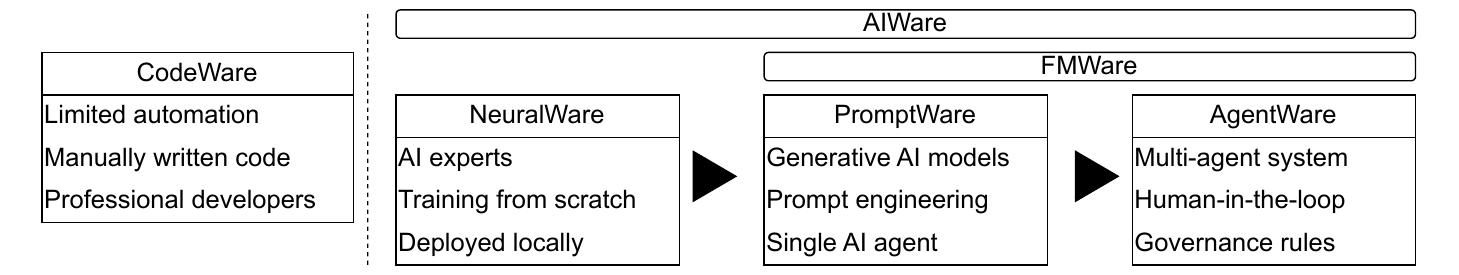}
	\caption{The evolution of software development paradigms   \citep{10.1145/3663529.3663820}.}
	\label{fig:aiware}
\end{figure}

Before the advent of AI models, the \textit{Codeware} paradigm has been widely adopted, involving professional programmers whose traditional lifecycle includes a well-established set of phases, \ie requirements, design, implementation, testing, deployment, and maintenance that are carried out mostly manually  \citep{10.1145/3663529.3663849}.
With the introduction of generative AI models, the paradigm has shifted from \textit{Codeware} to different instances of \textit{AIware} (see Figure \ref{fig:aiware}). In particular, the evolution of AIware has been divided into three main paradigms: \textit{NeuralWare}, \textit{PromptWare}, and \textit{AgentWare}. 

\textit{NeuralWare} development involves AI experts started to drive the development of a new generation of software \citep{8804457} by constructing a new set of assets such as datasets and models in a significantly more iterative and experiment-driven lifecycle. However, this new paradigm has brought new challenges to the software engineering community, \eg versioning of AI models \citep{10.5555/3618408.3619050}, AI-related code smells \citep{10.1145/3522664.3528620}, or additional requirements \citep{DEMARTINO2025107678}.

With the adoption and the further enhancements to the transformer architecture \cite{vaswani_attention_2017} and the launch of \CG in November 2022, the \textit{PromptWare} paradigm has emerged, as those models relies on natural language queries, knowns as \textit{prompts}, allowing the development by programmers without deep programming and AI skills, ultimately democratizing software creation. In particular, researchers have started to improve the prompting process by introducing new techniques, moving from zero-shot \citep{10.5555/3045118.3045347} to more advanced paradigms, \ie few-shot prompting \citep{LI2024112002} and chain-of-thought prompting \citep{yang_chain--thought_2023}. Although promising, PromptWare has also brought new challenges to the SE community, \eg hallucination \citep{liu_refining_2024,ji_survey_2023}, energy consumption \citep{10.1007/978-3-031-70245-7_12,castano_exploring_2023} and ethical concerns \citep{bhardwaj_investigating_2021}.


The last step of this phenomenon is presented by \textit{AgentWare} programming, that foster the interaction between different Foundation Model (FM) agents that are capable of decision-making, taking actions in an environment, and even interacting with other agents. In the SE domain, we can exemplify the usage of AgentWare in different phases of software development, with the majority of reported agents applied at the level of individual tasks, mainly in code generation and code quality assurance (e.g., static checking and testing), according to \cite{liu2024large} survey. However, we can also find AgentWare agents in end-to-end software development or maintenance, which shows that LLM-based agents can be utilized to solve more complicated tasks in SE.

One example of end-to-end software development is showcased by \cite{10.1145/3712003}, where two different games, Snake and Tetris, were developed using a platform called ChatDev \citep{qian2024chatdev}. Given that ChatDev organizes the development process into three phases (designing, coding, and testing), they employed five different agents: the CEO, CTO, programmer, reviewer, and tester. From this case study, the authors concluded that the process was cost-effective and efficient, as the total expenditure for generating these two games was relatively low per prompt submission attempt. However, when tasked with more complex projects like the Tetris game, the system struggled to generate correct code.

\cite{UMARZESHAN2026112792} proposed LAMPS, a multi-agent system using collaborative LLMs to detect malicious PyPI packages. The system addresses the growing threat of malicious code in open-source repositories that traditional rule-based tools often miss due to their inability to capture semantic patterns. The research demonstrates that distributed LLM reasoning effectively detects malicious code and highlights benefits of modular multi-agent designs in software supply chain security.

Although different approaches have been proposed, there is still the need to have a curated taxonomy to promote the effective adoption of AI agents.

\subsection{Surveys on LLM-based MAS frameworks}

This section discusses existing survey, taxonomy, and comparison studies on MAS frameworks. \cite{handler_taxonomy_2023} proposed a MAS taxonomy with a focus on autonomy and alignment with human users' goals. In particular, this work introduces a comprehensive multi-dimensional taxonomy that adheres to the 4+1 view model \cite{469759} of software architecture tailored focusing on agent composition and orchestration. On top of the proposed architectural model, a set of feature models have been depicted and discussed. To evaluate the expressiveness, seven different frameworks have been discussed using the proposed taxonomy. However, the taxonomy do not provide a comparison in terms of metrics, discussing qualitative and conceptual aspects. 
	\cite{10.1145/3712003} envisioned a research agenda for MAS in software engineering. After collecting relevant work with a systematic literature review (SLR), the authors developed two different games using MAS, \ie Snake and Tetris, to assess the performance of MAS in solving complex programming tasks. Afterward, they identify a set of challenges and future opportunities in the field, which have been categorized into two main dimensions, \ie enhancing individual agent capabilities and optimizing agent synergy. Concerning the single agents, it is crucial to refine the role-playing capabilities to build specialized agents by following three main steps, namely  \textit{(i)} identifying key SE roles using market analysis and involving stakeholder; \textit{(ii)} understanding the limitations of LLM-based agents such as hallucination and performance degradation; and \textit{(iii)} tailoring AI agents with specialized knowledge, prompts, and continuous learning. Concerning the optimization of the agents' synergy, the interaction with human expert and handling privacy concerns are the main challenges. 
	\cite{rasheed_autonomous_2025} proposed an initial approach to evaluate 12 different LLM agents by conducting two different experiments to support the Software Development Life Cycle (SDLC). In particular, the agents produced a requirements engineering specification, a software design plan, commented software code, a test and deployment plans for ten different software projects, spanning from snake game to a simple e-commerce application. The agents are then compared in terms of LOC and time taken to complete the tasks. 
	\cite{germanakos_conversational_2025} conducted a comprehensive comparison of existing Conversational Agents (CAs), spanning from neural network agents to cutting-edge transformers-based ones. First, a formal definition of CAs has given thorough the analysis of their dynamics in terms of domains of use and user experience. Then, the authors provide a categorization of the type of agents, \ie modular and end-to-end approaches, and their goals, \ie task-oriented and open-goal implementations. The conducted analysis showed that \textit{(i)} task-oriented CAs prioritize efficiency in task completion while open-goal agents focus on building relationships and delivering personalized interactions; and \textit{(ii)} a definitive solution to the challenges of automated conversation if far from to be definitive given the fast-evolving nature of the CAs. 

While prior studies have provided comprehensive analyses of LLM-based MAS, they have largely overlooked these frameworks from a developer's perspective--for example, in terms of documentation quality and ease of use. Moreover, although efficiency-related factors such as token consumption, number of requests, execution time, and cost are critical in practice, there is still a lack of systematic and comparative evaluation along these dimensions.  
Our work aims to fill this gap by providing a comprehensive taxonomy that has been elicited by analyzing both research papers and existing frameworks supporting the development of MAS. In addition, we provide a comparison of the frameworks, 
which can help developers to choose the most suitable framework for their needs.





\section{Qualitative Evaluation}
\label{sec:Results}


	We describe the methods and results related to our qualitative analysis. In particular, Section \ref{sec:features} presents the literature review and the functionalities available in MAS frameworks. Afterward, Sections~\ref{sec:RQ1} and~\ref{sec:RQ2} report and explain the obtained results to answer the following research questions:

\mybox{\textbf{\small{RQ$_1$}}}{gray!10}{gray!10}{To what extent do existing frameworks cover MAS foundational concepts?}

and 

\mybox{\textbf{\small{RQ$_2$}}}{gray!10}{gray!10}{What are the main characteristics of the considered MAS frameworks?}

\subsection{Literature Analysis} 
\label{sec:features}


To find suitable MAS frameworks and tools, we reviewed existing literature in the field of software engineering and code development by conducting a multi-vocal study \citep{neto_multivocal_2019}. Besides that, 
		we considered existing systematic studies, including both gray and white literature. The methodology is composed of three main steps, i.e., \textit{Planning}, \textit{Conducting} and \textit{Mapping}, following existing empirical standards \citep{kitchenham_systematic_2010}.

\paragraph{Planning.}The first step involves the definition of the query and the source of knowledge, \ie the selected digital library. In the scope of the paper, we consider the Scopus digital library\footnote{\url{https://www.scopus.com/}} as it is one of the largest and most comprehensive databases of peer-reviewed literature in the field of computer science. In addition, the platform enables searching for gray literature, \eg preprints and conference papers, that are not indexed in other digital libraries. Scopus covers preprints from 2017 onwards from arXiv, ChemRxiv, bioRxiv, medRxiv, SSRN, TechRxiv, and Research Square, thus granting a wide coverage of the most relevant preprint services.  While we acknowledge that the search is not exhaustive in terms of the identified papers, we aim to identify a ``quasi-gold'' set of academic works that can be used to elicit the foundational concepts of a MAS. Nevertheless, we adhere to formal guidelines proposed for conducting systematic studies that also involve gray literature \citep{garousi_guidelines_2019} without running additional steps, \eg snowballing.

\begin{table}[h!]
	\centering
	\caption{Keyword groups and final boolean search query.}
	\begin{tabular}{| c | p{9cm} |} \hline
		\textbf{Group} & \textbf{Keywords}  \\ \hline
		Group 1 & MAS, multi-agent system, LLM, large-language models, pre-trained language model  \\ \hline
		Group 2 & evidence-based software engineering survey, structured review, systematic review, literature review, literature analysis, in-depth survey, literature survey, meta-analysis, past studies, subject matter expert, analysis of research, empirical body of knowledge  \\ \hline
		Combined & (mas OR multi-agent AND system AND llm OR large-language AND models OR pre-trained AND language AND model AND ``evidence-based software engineering'' OR survey OR ``structured review'' OR ``systematic review'' OR ``literature review'' OR ``literature analysis'' OR ``in-depth survey'' OR ``literature survey'' OR ``meta-analysis'' OR ``past studies'' OR ``subject matter expert'' OR ``analysis of research'' OR ``empirical body of knowledge'')  \\ \hline
		
	\end{tabular}
	\label{tab:query}
\end{table}


The search string is composed of two different group of keywords, \ie MAS-related (Group 1) and Systematic studies (Group 2), that are eventually combined in the final boolean query as shown in Table \ref{tab:query}. We use the \textit{TITLE-ABS-KEY} field to search for the keywords in the title, abstract, and keywords of the papers. We also limited the search to the Computer Science domain and to articles published in well-ranked venues according to the CORE ranking\footnote{\url{https://portal.core.edu.au/conf-ranks/}} and Scimago Journal Rank (SJR).\footnote{\url{https://www.scimagojr.com/}} To find relevant studies, we followed the guidelines proposed by existing research \citep{kitchenham_systematic_2010} for conducting \textit{tertiary studies} in which the subject of the study are systematic reviews. While 
	the search string is not exhaustive, it represents a starting point to elicit pivotal components of a MAS system.

\begin{table}[h!]
	\centering
	\scriptsize
	\caption{Inclusion and exclusion criteria for the literature analysis.}
	\begin{tabular}{|p{9.5cm}|}
		\hline
		\multicolumn{1}{|c|}{\textbf{Inclusion criteria}} \\ \hline 
		\tick~Surveys, literature reviews, or similar that focus on MAS based on LLMs and/or pre-trained models.     
		\\ \hline 
		\tick~Articles that provide governance rules, guidelines, and other foundational concepts for building MAS.   
		\\ \hline 
		\tick~Conference and journal paper published in well-ranked venues according to CORE ranking and Scimago Journal Rank (SJR). In addition, we considered relevant papers that have not been published but available on preprint services, \eg arXiv. 
		\\\hline
		\tick~Articles written in English and published in the Computer Science domain.
		\\ \hline
		\tick~Studies published from January 2023 to January 2025
		\\ \hline
		\multicolumn{1}{|c|}{\textbf{Exclusion criteria}} \\ \hline 
		\untick~Studies that focus on multi-agent systems but not large language models. \\ \hline
		\untick~Papers that propose a new MAS but do not provide a systematic study or comparison of them. \\ \hline
		\untick~Workshops, vision/short papers, and posters. \\ \hline  
		
	\end{tabular}
	\label{tab:criteria_white}
\end{table}

We then defined a set of inclusion and exclusion criteria to filter the results, summarized in Table \ref{tab:criteria_white}. In particular, we did not consider papers that propose a new MAS. Instead, we focused on studies that provide foundational concepts for building MAS. In addition, we imposed the following criteria: 
	\emph{(i)} the search string is composed of keywords related to MAS and LLMs; \emph{(ii)} the search was limited to Computer Science; \emph{(iii)} only articles published in premier 
	venues according to CORE ranking and Scimago Journal Rank (SJR) were selected; \emph{(iv)} the articles must be written in English; and finally \emph{(v)} the search was confined to the most 2 recent years, \ie from January 2023 to January 2025.

\begin{table}[h!]
	\centering
	\scriptsize
	\caption{Inclusion and exclusion criteria for MAS frameworks.}
	\begin{tabular}{|p{9.5cm}|}
		\hline
		\multicolumn{1}{|c|}{\textbf{Inclusion criteria}} \\ \hline 
		\tick~Frameworks with supporting \GH repositories.
		\\ \hline 
		\tick~Frameworks that provide proper documentation, tutorials, or similar to extract foundational concepts of MAS.
		\\\hline
		\tick~Open-source frameworks or those that provide permissive licenses.
		\\ \hline
		\multicolumn{1}{|c|}{\textbf{Exclusion criteria}} \\ \hline 
		\untick~Frameworks to develop single LLM-based systems or that do not allow orchestration among agents. \\ \hline
		\untick~Frameworks that have fewer than 10,000 \GH stars. \\ 
		\hline  
	\end{tabular}
	\label{tab:criteria_tools}
\end{table}

We used \GH as the base for selecting frameworks, 
	as it provides a wide range of open-source projects. The inclusion and exclusion criteria 
	applied 
	are summarized in Table \ref{tab:criteria_tools}. We first identified a \GH awesome list\footnote{\url{https://github.com/kaushikb11/awesome-llm-agents}} of tools, then we considered open-source frameworks that provide proper documentation and/or a supporting \GH repository. Tools with fewer than 10,000 stars were excluded. While the number of stars might not be sufficient to judge the quality of the frameworks, it can be used together with the number of forks 
	to evaluate the popularity of the tool \citep{BORGES2018112}. Moreover, we checked tech blog posts, forums, and other sources to identify additional relevant tools and frameworks falling within the definition of gray literature provided by \citep{garousi_guidelines_2019}. For the literature analysis, we independently analyzed the tools and then discussed the results to reach a consensus on the final set. To facilitate future research, we specified the selection process in the replication package available online \citep{replicationPackage}.

\begin{table}[t!]
	\centering
	\small
	\scriptsize
	\caption{Functionalities of MAS frameworks.}  
	\begin{tabular}{|p{1.0cm}|p{2.8cm}|p{6.8cm}|}    \hline  
		\textbf{Alias} & \textbf{Name} & \textbf{Description}    \\ \hline 
		\textbf{F1} & MAS core architecture &   We analyze if the tool allows for the definition of the core features of a multi-agent system identified in the literature, \ie definition of agents, orchestration, concept of memory and external tools as defined in existing studies \citep{10.1145/3712003,xi_rise_2025}. \\ \hline
		
		\textbf{F2} & MAS type & This concerns whether the tool enables the integration of custom agents or agents coming from a different architecture, 
		\eg AI models, transformers. If this feature is not provided, we consider the system as Homogeneous, otherwise it is Heterogeneous. \\ \hline
		
		\textbf{F3} & Role specification & With this feature, we check if the platform allows us to define the role of an agent, \eg PromptAgent, SupervisorAgent, or other roles. \\ \hline
		
		\textbf{F4} & Tool support & This dictates if the tool allows for the usage of external tools, \eg APIs, or other tools. \\ \hline
		
		\textbf{F5} & Remote access to agents & We analyze if the tool facilitates remote access to agents, \eg through a Web interface or other means, or if the agents are only available locally. \\ \hline
		
		\textbf{F6} & Agent monitoring & This evaluates whether the platform supports any monitoring facility, \eg evaluation metrics, or telemetry functions.	\\ \hline
		
		\textbf{F7} &  Human feedback integration & We analyze if it is possible to embed human feedback in the agent's decision making process. Note that chat is not considered as human feedback. \\ \hline	
		
		\textbf{F8} & Agent comparison & This is related to the functionality to allows one to compare agents, \eg through a web interface or other means. \\ \hline		
		
		\textbf{F9} & Reusage of benchmarks datasets & Platforms could enable the reuse of benchmark datasets for the evaluation of the agents through a specific interface or API.  \\ \hline		
		
		\textbf{F10} & Discovery capabilities & Discovering new agents or new functionalities, \eg store or marketplace for agents or tools is a useful functionality, and we check if it is enabled by the considered platforms. \\ \hline		
		
		
	\end{tabular}	
	\label{tab:features}
\end{table}

\paragraph{Conducting.} Once we defined the search string and the inclusion and exclusion criteria, we obtained a total number of \tot papers that match the identified keywords. Then, by applying the defined criteria, we ended up with \final relevant papers. Five of them are classified as \textit{white literature} as they have peer-reviewed in high-ranked journals and conferences, while the remaining papers fall within the \textit{gray literature} category, \ie being available on preprint services. 
	Similarly, we started from a pool of 20 different frameworks that are publicly available on \GH. After the application of the criteria, we excluded two frameworks with fewer than 10,000 stars, and three with subscriptions. Finally, we got a final set of 16 frameworks being relevant to the scope of the paper.

\paragraph{Mapping.} The final step of the methodology involves mapping the selected papers and frameworks to the identified dimensions. We first read the full text of each paper to extract an initial list of features for each dimension. Afterward, we discussed the results to reach a consensus on the final list of features, and mapped the features to the corresponding papers. Eventually, we obtained 
	the foundation features summarized in Table \ref{tab:features}.

\begin{table}[h!]
	
	\centering
	\small
	\scriptsize
	\caption{Characteristics.}  
	\begin{tabular}{|p{1.0cm}|p{2.8cm}|p{6.8cm}|}    \hline  
		\textbf{Alias} & \textbf{Characteristic} & \textbf{Description}   \\ \hline 
		\textbf{C1} & Installation & It describes how to install the tool, including any dependencies or prerequisites that need to be met. There are also instructions for setting up the tool in a local environment, as well as any necessary configurations.  
		\\ \hline
		
		\textbf{C2} & Developer Interface & This concerns the 
		ease of use and intuitiveness of the supported interface. It is also related to the quality of the documentation provided, including its clarity, completeness, and organization. This includes evaluating the availability of tutorials, examples, and API references. \\ \hline
		\textbf{C3} & Model and Tools Integration & Support for integrating with different AI models, including LLMs and other types of models, as well as external services. We assess the flexibility of the tool in terms of model selection and integration, as well as its ability to work with different model architectures. This includes evaluating the availability of pre-built integrations and the flexibility of the tool in terms of custom integration. \\ \hline
		
		\textbf{C4} & Agent Creation & Support for creating and managing agents, \ie the ease of creating new agents, including the availability of templates and examples. In addition, this concerns how role and external knowledge are defined in the tool.  \\ \hline
		
		\textbf{C5} & Agent Orchestration &  Support for orchestrating agents and managing their interactions, \ie the ease of defining workflows and coordinating multiple agents, including the availability of pre-built orchestration patterns.  \\ \hline
		
		\textbf{C6} & Monitoring &  Support for monitoring and debugging agents, \ie the availability of monitoring tools, including logging and telemetry features. This includes evaluating the ease of tracking agent performance and identifying issues during development and deployment. \\ \hline		
	\end{tabular}	
	\label{tab:Characteristics}
\end{table}

\vspace{.1cm}
\noindent
\hand \textbf{Addressing RQ$_1$.} 
	To report the main advantages and disadvantages of each tool from a developer's perspective, we defined the technical characteristics 
	shown in Table~\ref{tab:Characteristics}. On one hand, they cover technical aspects related to the installation and integration of the tools, \eg installation process, model and tools integration, and agent creation. On the other hand, they partially overlap with the foundation functionalities discussed in Table \ref{tab:features}. This aims 
	to provide both practical and foundational evidence to support the development of MAS.

\vspace{.1cm}
\noindent
\hand \textbf{Addressing RQ$_2$.} The search in the literature analysis resulted in several frameworks, and to answer \textbf{RQ$_2$} we had to narrow down the corpus to focus only on the most representative 
	ones. To this end, the first two authors of this paper independently analyzed the tools identified in \textbf{RQ$_1$}, 
	picking three different frameworks for each programming paradigm, \ie low and high code. Moreover, we chose the frameworks that have the highest number of \GH stars and forks, as well as maintain a trade-off between the number of features provided and the ease of use. The analysis led to the identification of \emph{eight} representative tools, 
	\ie AutoGen \citep{microsoft-autogen}, AutoGPT \citep{autonomous-gpt}, Dify \citep{dify}, Flowise \citep{flowise}, Haystack \citep{haystack}, Llama Index \citep{llama-index}, OpenAISDK \citep{openagents}, and Semantic Kernel \citep{microsoft-semantic-kernel}. For the sake of presentation, in the rest of this paper, we report and analyze the tools in alphabetical order, without grouping them into two distinct categories, \ie low code and high code. 
	

	\subsection{Coverage of Foundational Concepts}
	\label{sec:RQ1}

	Table \ref{tab:tools_comparison} provides an overview of the selected MAS frameworks and their support for the ten features defined in Section \ref{sec:features}. We examined the official documentation provided on GitHub or the framework’s website and similar documents to indicate if a feature is fully supported \CIRCLE, partially supported \LEFTcircle, or not supported \Circle. A feature is considered fully supported when the framework provides all the characteristics in Table \ref{tab:features}. A feature is partially supported when the framework implements some of the characteristics but lacks others. Finally, a feature is not supported when none of the characteristics described for that feature is present in the framework. The classification of each framework was performed by a single author of the present work, with the classification distributed among the authors.
	
	In addition, we classify the tools into two categories \ie \textit{low-code} and \textit{high-code}. The former characterizes frameworks that facilitate the MAS development by providing graphical capabilities for newcomer developers. The latter represents traditional frameworks that require developers to write code to implement the MAS even though they may provide some built-in functions and classes to facilitate the development.

	\begin{table}[h!]
\scriptsize
\centering
\caption{Overview of MAS frameworks grouped by development paradigm.}
\label{tab:tools_comparison}
\begin{tabular}{l c c c c c c c c c c}
	\toprule
	\textbf{Name} 
	& \textbf{F1} & \textbf{F2} & \textbf{F3} & \textbf{F4} & \textbf{F5} & \textbf{F6} & \textbf{F7} & \textbf{F8} & \textbf{F9} & \textbf{F10} \\
	\midrule
	\multicolumn{11}{l}{\textbf{Low-code}} \\
	\midrule
	\rowcolor{gray!20}
	Haystack \citep{haystack}
	& \CIRCLE & \CIRCLE & \CIRCLE & \CIRCLE & \CIRCLE & \CIRCLE & \Circle & \CIRCLE & \LEFTcircle & \Circle \\
	\rowcolor{gray!20}
	Dify \citep{dify}
	& \CIRCLE & \CIRCLE & \CIRCLE & \CIRCLE & \CIRCLE & \CIRCLE & \CIRCLE & \CIRCLE & \CIRCLE & \CIRCLE \\
	\rowcolor{gray!20}
	Flowise \citep{flowise}
	& \CIRCLE & \CIRCLE & \CIRCLE & \CIRCLE & \CIRCLE & \LEFTcircle & \Circle & \LEFTcircle & \Circle & \CIRCLE \\
	\rowcolor{gray!20}
	AutoGPT \citep{autonomous-gpt}
	& \CIRCLE & \CIRCLE & \CIRCLE & \CIRCLE & \CIRCLE & \LEFTcircle & \Circle & \Circle & \Circle & \CIRCLE \\
	Botpress \citep{botpress}
	& \LEFTcircle & \Circle & \CIRCLE & \CIRCLE & \CIRCLE & \LEFTcircle & \CIRCLE & \Circle & \Circle & \LEFTcircle \\

	\midrule
	\multicolumn{11}{l}{\textbf{High-code}} \\
	\midrule
	\rowcolor{gray!20}
	AutoGen \citep{microsoft-autogen}
	& \CIRCLE & \CIRCLE & \CIRCLE & \CIRCLE & \LEFTcircle & \CIRCLE & \CIRCLE & \Circle & \CIRCLE & \Circle \\
	\rowcolor{gray!20}
	Llama Index \citep{llama-index}
	& \CIRCLE & \CIRCLE & \CIRCLE & \CIRCLE & \Circle & \CIRCLE & \CIRCLE & \CIRCLE & \Circle & \Circle \\
	\rowcolor{gray!20}
	OpenAISDK \citep{openagents}
	& \CIRCLE & \CIRCLE & \CIRCLE & \CIRCLE & \Circle & \LEFTcircle & \Circle & \LEFTcircle & \Circle & \CIRCLE \\
	\rowcolor{gray!20}
	Semantic Kernel \citep{microsoft-semantic-kernel}
	& \CIRCLE & \CIRCLE & \CIRCLE & \CIRCLE & \CIRCLE & \CIRCLE & \CIRCLE & \Circle & \Circle & \Circle \\
	\rowcolor{gray!60}
	LangChain \citep{langchain}
	& \CIRCLE & \CIRCLE & \CIRCLE & \CIRCLE & \CIRCLE & \LEFTcircle & \CIRCLE & \CIRCLE & \LEFTcircle & \CIRCLE \\
	Camel \citep{camel}
	& \CIRCLE & \CIRCLE & \CIRCLE & \CIRCLE & \CIRCLE & \Circle & \Circle & \Circle & \Circle & \Circle \\
	CrewAi\citep{crewai}
	& \CIRCLE & \CIRCLE & \CIRCLE & \CIRCLE & \CIRCLE & \Circle & \CIRCLE & \Circle & \Circle & \Circle \\   
	ix \citep{ix}
	& \CIRCLE & \CIRCLE & \CIRCLE & \CIRCLE & \CIRCLE & \Circle & \Circle & \Circle & \Circle & \Circle \\

	MetaGPT \citep{metagpt}
	& \LEFTcircle & \CIRCLE & \CIRCLE & \CIRCLE & \Circle & \Circle & \LEFTcircle & \Circle & \Circle & \Circle \\
	
	Smolagents \citep{smolagents}
	& \CIRCLE & \CIRCLE & \CIRCLE & \CIRCLE & \CIRCLE & \CIRCLE & \Circle & \LEFTcircle & \Circle & \Circle \\
	Agno \citep{agno}
	& \CIRCLE & \CIRCLE & \CIRCLE & \CIRCLE & \CIRCLE & \Circle & \Circle & \Circle & \Circle & \CIRCLE \\
	\bottomrule
\end{tabular}
\end{table}

From the analysis, we found that the most complete framework is LangChain~\citep{langchain}, as it provides support (full and partial) for all the elicited functionalities. This is quite expected as it was one of the first MAS frameworks to be developed, and it has been continuously improved since its release. Similarly, Dify~\citep{dify} supports all the identified functionalities except for F6, \ie Agent Monitoring. Contrariwise, ix~\citep{ix} and Camel \citep{camel} support the first five functionalities, neglecting the remaining ones. Concerning the framework accessibility, we report that Amazon Bedrock, VertexAI and Botpress require a subscription to use their services, while the other frameworks are open-source and can be used for free. Therefore, we excluded these three frameworks from our further analyses.

Overall, the low-code frameworks covers more functionalities compared to high-code, especially qualitative aspects like benchmarking, monitoring, and evaluation. On the one hand, low-code frameworks are designed to be more user-friendly and accessible, 
often providing built-in tools. On the other hand, high-code frameworks focus more on flexibility and extensibility, leaving qualitative aspects to be implemented by developers as needed.

From the initial list, we further refined the list of tools by considering the following aspects: \emph{(i)} they offer a permissive license, \eg developers can reuse the tool without any subscription or commercial license; \emph{(ii)} they have a large number of stars and forks, \ie they are well-maintained and have a vibrant community of developers; 
and \emph{(iii)} they provide a comprehensive documentation and examples to support developers in using the tool. By applying these criteria, we selected eight (marked in gray in Table \ref{tab:tools_comparison}) frameworks for our analysis and 
evaluated them using the \RM summarization task, \ie the \ME scenario~\citep{10.1145/3696630.3728511} presented in Section~\ref{sec:quantitative}. In such a way, we covered all the ten functionalities using different frameworks, thus providing a comprehensive analysis of the MAS frameworks.

\begin{framed}
\noindent \textbf{Answer to $RQ_1$:} Concerning low-code platforms, Dify \citep{dify} is the one that covers most of the considered functionalities, \ie 9 over 10, and both Haystack \citep{haystack} and Flowise \citep{flowise} come at the second place, making available 7 over 10 functionalities. With respect to high-code platforms, 
	LangChain \citep{langchain} is the most complete one, with more than 8 fully supported, and 2 partially supported  functionalities. Still, there are some gaps in terms of agent monitoring and evaluation, which are not fully supported by a few frameworks. Essentially, the low-code platforms, \ie Haystack \citep{haystack}, Flowise \citep{flowise}, AutoGPT \citep{autonomous-gpt}, Botpress \citep{botpress}, tend to cover more qualitative aspects compared to high-code ones, even though the integration between the two paradigms is supported. 
	\end{framed}

	\subsection{Characteristics of the Frameworks}
	\label{sec:RQ2}

	
	We investigate different aspects of the frameworks to answer the following research question. In particular, the first three authors implemented the \ME scenario independently and received the same training and documentation materials. Then, the fourth author aggregates the results to reach a consensus on the main characteristics of the frameworks. Our aim is to answer the following research question:
	
	\mybox{\textbf{\small{RQ$_2$}}}{gray!10}{gray!10}{What are the main characteristics of the considered MAS frameworks?}
	

	To this end, we consider the characteristics defined in Table~\ref{tab:Characteristics}, inspecting the frameworks to understand how well they cover them. 
	
	\subsubsection{C1 - Installation}
	
	Table \ref{tab:Installation} summarizes the installation process for each framework, highlighting the setup method and the quickstart support in terms of documentation, tutorial, or similar materials. 
	Overall, all the frameworks provide well-structured documentation, even though at different levels of detail. In terms of installation, high-code frameworks can be installed using common package managers, \eg pip or npm, while low-code ones require Docker to be run, especially the one classified as low-code, possibly increasing the complexity of the setup for newcomers. In addition, some frameworks provide a web-based dashboard to facilitate the development and management of MAS applications. 

	\begin{table}[h!]

\centering		
\caption{Installation.}
\scriptsize
\begin{tabular}{|l|c|c|c|p{3.3cm}|}    \hline
	\textbf{Framework} & \textbf{Local} & \textbf{Cloud} & \textbf{Docker} & \textbf{Quickstart support} \\ \hline 
	AutoGen \citep{microsoft-autogen} & \tick & \untick & \untick & Documentation, Tutorials \\ \hline
	AutoGPT \citep{autonomous-gpt} & \tick & \untick & \tick & Documentation, Tutorials \\ \hline
	Dify \citep{dify} & \tick & \tick & \tick & Documentation, Templates  \\ \hline 
	Flowise \citep{flowise} & \tick & \untick & \tick & Tutorials \\ \hline
	Haystack \citep{haystack} & \tick & \untick & \untick & Documentation\\ \hline
	Llama Index \citep{llama-index} & \tick & \untick & \untick & Visualization \\ \hline
	OpenAISDK \citep{openagents} & \tick & \untick & \untick & Documentation \\ \hline
	Semantic Kernel \citep{microsoft-semantic-kernel} & \tick & \untick & \untick & Notebooks, Examples \\ \hline
\end{tabular}	
\label{tab:Installation}
\end{table}



	\smallskip
	\noindent
	$\triangleright$ \textbf{AutoGen}. The setup is straightforward, using common \texttt{pip} commands which will install the AgentChat and Extension packages. 
	The documentation provides simple examples of developing agents, such as agents counting numbers or providing weather information. 

	\smallskip
	\noindent
	$\triangleright$ \textbf{AutoGPT}. The tool can be installed into two ways, \ie locally or the cloud. However, the cloud service is not yet available to all users since the tool is still in the beta version. As a result, the tested version was run locally using Docker for the backend and \texttt{npm} (Node Package Manager) for the front end. 
	
	\smallskip
	\noindent
	$\triangleright$ \textbf{Dify}. Similarly to AutoGPT, Dify can be run locally using Docker, or 
	deployed in the cloud with either a free plan or a paid option. On the one hand, the cloud tool requires minimal setup, it has some limitations on the number of apps, requests, and memory. On the other hand, the self-hosted version is free but requiring more effort in the setup. 
	
	\smallskip
	\noindent
	$\triangleright$ \textbf{Flowise}. The framework requires familiarity with JavaScript development, particularly the use of \texttt{npm} using the standard installation process. This method involves setting up the local environment manually, which may pose a barrier for beginners or people with no experience with JS technologies. 

	\smallskip
	\noindent
	$\triangleright$ \textbf{Haystack}. Developers can choose to install Haystack using \textit{pip} commands or relying on deepset studio,\footnote{\url{https://www.deepset.ai/deepset-studio}} a tailored IDE, even though it needs a dedicated account. Besides the technical documentation that offers a good overview of basic building blocks, a set of tutorials is available on a dedicated \GH repository.\footnote{\url{https://github.com/deepset-ai/haystack-tutorials}} 
	
	\smallskip
	\noindent
	$\triangleright$ \textbf{LlamaIndex}. The installation is simple and streamlined, with support for dependency management systems such as Conda and Poetry. For most use cases, a single command--pip install llama-index--is sufficient for enabling core functionalities, including the definition of structured workflows and agent-based architectures. 

	\smallskip
	\noindent
	$\triangleright$ \textbf{OpenAI SDK}. The installation process is straightforward, requiring only the installation of the SDK package via \texttt{pip}. Additional setup includes configuring environment variables used by agents and tools. In addition, the SDK offers a quickStart guide that effectively walks users through installation and basic usage. It covers how to define agents, add guardrails, and orchestrate workflows, making it easy for new users to get started.

	\smallskip
	\noindent
	$\triangleright$ \textbf{Semantic Kernel}. The platform supports three different languages, \ie C\#, Python, and Java,  with instructions for the corresponding package managers. 
	Alternatively, developers can directly pass parameters to the instantiated function responsible for setting up models. 

	\subsubsection{C2 - Developer Interface}
	
	Table \ref{tab:DeveloperInterface} summarizes the characteristics related to developer interface for each framework, highlighting the CLI, IDE/SDK, and web-based dashboard support for each tool. The CLI column indicates whether the framework provides a command-line interface for developers to interact with the system, while the IDE/SDK column signifies the presence of 
	an integrated development environment or software development kit. Finally, the web-based dashboard column indicates if the framework provides a web-based interface for managing and monitoring the MAS applications.

\begin{table}[h!]	
\centering
\scriptsize
\caption{Developer Interface.}  
\begin{tabular}{|l|c|c|c|}    \hline  
	\textbf{Framework} & \textbf{CLI} & \textbf{IDE/SDK} & \textbf{Web-based dashboard}  \\ \hline 
	AutoGen \citep{microsoft-autogen} & \tick & \tick & \untick \\ \hline
	AutoGPT \citep{autonomous-gpt} & \tick & \untick & \tick \\ \hline
	Dify \citep{dify} & \untick &  \tick & \tick 	 \\ \hline
	Flowise \citep{flowise} & \untick & \untick & \tick \\ \hline
	Haystack \citep{haystack} & \tick & \untick & \untick  \\ \hline
	Llama Index \citep{llama-index} & \tick & \untick & \untick  \\ \hline
	OpenAISDK \citep{openagents} & \tick & \tick & \untick \\ \hline	
	Semantic Kernel \citep{microsoft-semantic-kernel} & \tick & \untick &  	\untick	
	\\ \hline
	
\end{tabular}	
\label{tab:DeveloperInterface}
\end{table}


\smallskip
\noindent
$\triangleright$ \textbf{AutoGen}. All the implementation is carried out by writing code using Python and users can opt for using Python scripts or Jupyter notebooks. The documentation is easy to follow and covers different aspects of the tool. Additionally, it provides simple examples to be reimplemented by the user.

\smallskip
\noindent
$\triangleright$ \textbf{AutoGPT.} Users can interact with the system through a dashboard accessed via a web application that provides three different views, \ie Build, Library, and Marketplace. In the Build section, users can create workflows using a system of visual blocks to perform specific tasks with LLM agents while the Library view store and display the created workflows. 

\smallskip
\noindent
$\triangleright$ \textbf{Dify}. It offers a user interface for the system, designed as an intuitive dashboard, enabling users to seamlessly 
create and manage various applications, \eg chatbots, agents, or workflows. This dashboard also allows for the modification of the system's configuration and models' setup. 

\smallskip
\noindent
$\triangleright$ \textbf{Flowise}. The platform provides a browser-based interface characterized by its drag-and-drop interaction model. Users can visually construct workflows by selecting components from a categorized library and positioning them on a design canvas. Each component is 
configured to suit specific functional needs. 

\smallskip
\noindent
$\triangleright$ \textbf{Haystack}. The open-source version does not include a native user interface, and it is primarily code-driven, making it ideal for developers familiar with Python. Meanwhile, deepset Studio provides a drag-and-drop editor to build and manage pipelines visually. In the scope of our paper, we evaluated the open-source version of the interface, \ie without relying on the IDE. 

\smallskip
\noindent
$\triangleright$ \textbf{LlamaIndex}. It relies on Python classes, decorators, and agent-based patterns, with type-safe constructs often defined via Pydantic. It does not provide a graphical or declarative interface for composing workflows, as all components must be implemented directly in code. 

\smallskip
\noindent
$\triangleright$ \textbf{OpenAI SDK}. 
The framework offers both the integration with native Python and SDK. The former consists of interfaces and classes for orchestrating and chaining agents, including decorators and inheritance capabilities. 

\smallskip
\noindent
$\triangleright$ \textbf{Semantic Kernel}. Similar to LlamaIndex, Semantic Kernel does not provide graphical features but the official version has comprehensive documentation to outline its functionalities and examples to be used as Python scripts. Nevertheless, the same outcome can be achieved through various functions that can be somewhat confusing for newcomers. 


\subsubsection{C3 - Agents and Tools Integration}

Table \ref{tab:ToolsIntegration} depicts an overview of the integration strategies for agents and external tools. The function calling column indicates whether the framework supports invoking third-party APIs, while the hosted tools column dictates the presence of 
pre-defined tools that can be used within the MAS application. The agent-as-a-tool column signals the availability of utilities 
to incorporate agents 
as reusable components within workflows. 
The last column shows the presence of 
a marketplace for sharing and discovering agents and tools.

\begin{table}[h!]		
\centering
\scriptsize
\caption{Agents and tools integration.}  
\begin{tabular}{|l|c|c|c|c|}    \hline  
	\textbf{Platform} &   \textbf{Func. call.} & \textbf{Hst. tools}& \textbf{Agent-as-Tool} &\textbf{Mrkt. place} \\ \hline 
	AutoGen \citep{microsoft-autogen} & \tick & \tick & \untick & \untick \\ \hline
	AutoGPT \citep{autonomous-gpt}  & \tick & \tick & \untick & \tick \\ \hline
	Dify \citep{dify}  & \untick & \tick & \untick & \tick \\ \hline
	Flowise \citep{flowise} & \tick & \tick & \untick & \untick \\ \hline
	Haystack \citep{haystack} & \tick & \tick & \untick & \untick \\ \hline
	Llama Index \citep{llama-index} & \tick & \tick & \untick & \tick \\ \hline	
	OpenAISDK \citep{openagents} & \tick & \tick & \tick & \untick \\ \hline
	Semantic Kernel \citep{microsoft-semantic-kernel} & \tick & \untick & \untick & \untick \\ \hline			
\end{tabular}	
\label{tab:ToolsIntegration}
\end{table}

\smallskip
\noindent
$\triangleright$ \textbf{AutoGen}. Agents 
can be utilized by installing the extension specific to the provider used even though some of them are still in the experimental phase, \eg Anthropic and Ollama. During configuration, the user can choose which model the agent will utilize. During runtime, it is possible to switch the model as long as the user has written the code for doing this at execution time. 

\smallskip
\noindent
$\triangleright$ \textbf{AutoGPT.} It offers the possibility to use different agents from different providers by specifying the API key. 
However, the chosen agent cannot be changed at runtime as it is linked to the execution node. 

\smallskip
\noindent
$\triangleright$ \textbf{Dify}. It supports various models within the workflow application type. To access different types of agents in the application, users must install the corresponding model plugin, available in the application marketplace. Once installed, they can be accessed within the app. Additionally, users can easily set up and switch agents in the application via the dashboard is straightforward while configuring the workflow.

\smallskip
\noindent
$\triangleright$ \textbf{Flowise}. The platform supports integration with a wide range of agents providers, offering developers flexibility in choosing and switching between LLMs. Changing from one model to another is relatively straightforward, as it only requires replacing or reconfiguring the appropriate model component within a given flow. 

\smallskip
\noindent
$\triangleright$ \textbf{Haystack}. It provides a dedicated functionality to allow for agents integration, called ``Generators,'' supporting both notable providers and local models, making the integration straightforward. Similarly, it integrates external tools like retrievers, fetchers, and other components developed by the Haystack team or community-based are integrated into pipelines. 

\smallskip
\noindent
$\triangleright$ \textbf{LlamaIndex}. The integration of the model into LlamaIndex is designed to be highly adaptable, with native support for both remote LLMs such as OpenAI and local deployments through Ollama, OpenLLM, or vLLM. Due to its decoupled architecture, switching between agents typically requires only minor changes to the configuration or import paths, leaving the core logic of workflows untouched. 


\smallskip
\noindent
$\triangleright$ \textbf{OpenAI SDK}. While the framework supports OpenAI models natively, custom model classes must be implemented for external providers according to the required interface. This has an impact also on agents selection, \ie  third-party provides needs a corresponding class to be integrated in the workflows. OpenAI SDK provides built-in support for several tools, including Web search, file search, and a computing tool for automating tasks. 

\smallskip
\noindent
$\triangleright$ \textbf{Semantic Kernel}. Using and selecting different agents is straightforward even though the process require some effort for developers, as utilizing different model providers involves invoking a distinct function from the SDK. Therefore, changing models entails rewriting portions of the code. Additionally, users can incorporate multiple models into the kernel and can switch between them at runtime, as long as the code clearly specifies the desired method for doing so. 


\subsubsection{C4 - Agent Creation}

Table \ref{tab:AgentCreation} summarizes the agent creation process for each framework, highlighting the role specification, memory management, and template support. The role specification indicates how the agent's behavior is defined, while memory refers to the type of memory used by the agent, \ie short-term (ST) or long-term (LT). Finally, the Template column signals whether the framework provides ready-to-use designs for creating agents.

\begin{table}[h!]		
\centering
\scriptsize
\caption{Agent Creation.}  
\begin{tabular}{|l|c|c|c|}    \hline  
\textbf{Platform} & \textbf{Role specification} & \textbf{Memory} & \textbf{Template} \\ \hline 
AutoGen \citep{microsoft-autogen} & Class-based & ST, LT & \untick \\ \hline
AutoGPT \citep{autonomous-gpt} & Prompt-based & ST, LT & \untick \\ \hline
Dify \citep{dify} & Node-based  & ST &  \untick	 \\ \hline
Flowise \citep{flowise} & Prompt-based & ST & \untick \\ \hline
Haystack \citep{haystack} & Prompt-based & ST, LT & \untick \\ \hline
Llama Index \citep{llama-index} & Prompt-based  & ST, LT  & \tick \\ \hline				
OpenAISDK \citep{openagents} & Class-based  & ST, LT   & \untick \\ \hline	
Semantic Kernel \citep{microsoft-semantic-kernel} & File-based  & ST, LT   & \tick\\ \hline		

\end{tabular}	
\label{tab:AgentCreation}
\end{table}

\smallskip
\noindent
$\triangleright$ \textbf{AutoGen}. 
The creation process involves instantiating the class associated with the desired agent, including setting up the messages protocol. Therefore, the framework does not utilize templates for implementing agents. 
The usage of ST memory is enabled by utilizing code variables and storing the agents' last conversation with built-in functions. Additionally, it is possible to save the agent's state to disk for later retrieval. Instead, LT memory can be achieved by using third-party vector databases that are accessible during code execution.

\smallskip
\noindent
$\triangleright$ \textbf{AutoGPT}. Agents can be defined using any block under the AI category but no predefined template is provided. Once the type of block is chosen, the user can manually define the prompt to specify the behaviors and roles. 
Another option is to utilize predefined AI blocks, which implements agents specialized in well-known  tasks, \eg summarization or image generation. 
Concerning the memory, AutoGPT supports both ST and LT memory. 

\smallskip
\noindent
$\triangleright$ \textbf{Dify}. Agent creation is done by adding the LLM or Agent node to an existing workflow, even though roles and behaviours can be customized using prompts. In the workflow, the agents' ST memory essentially consists of its output, which can be accessed by subsequent nodes. Furthermore, LT memory can be built by using miscellaneous data sources \eg text files, Notion DB,\footnote{\url{https://www.notion.com/help/intro-to-databases}} or external websites. It is worth noting that Dify enables the creation of prompts where variables can be defined, offering greater flexibility for the prompts.

\smallskip
\noindent
$\triangleright$ \textbf{Flowise}.
The system does not provide templates for agent creation, but it allows the user to define the agent's role and behavior using prompts. However, the documentation does not offer any examples on how to implement agents, which may make it challenging for new users to understand how to create them effectively.

\smallskip
\noindent
$\triangleright$ \textbf{Haystack}.
Developers can define intelligent LLM-powered agents that interact with tools and manage internal state across multiple steps by exploiting the built-in Agent component. Each agent is configured with a chat generator, a set of tools, and specific exit conditions to control when the agent should stop. A system prompt can be added to define the agent's role or behavior, while a state schema enables ST memory management. 

\smallskip
\noindent
$\triangleright$ \textbf{LlamaIndex}. LlamaIndex provides abstractions for building LLM-based agents using two primary templates \ie FunctionAgent and ReActAgent. The FunctionAgent is designed for scenarios where LLM supports function calling, \ie invoking tools asynchronously and maintaining a history of interactions. The agent processes tool results, which are returned as messages with the role ``tool,'' and updates its context based on the outcomes. 


\smallskip
\noindent
$\triangleright$ \textbf{OpenAI SDK}.
Creating an agent involves instantiating the \texttt{Agent} class with the chosen model and instruction prompt. The base agent class can be extended to define custom behaviors and capabilities. Agent roles and behaviors are defined using the \texttt{instruction} parameter during initialization, which acts as the system prompt guiding the agent’s operation. 

\smallskip
\noindent
$\triangleright$ \textbf{Semantic Kernel}.
While the system does not provide predefined agents, Semantic Kernel drives the developers using YAML templates to facilitate the agent instructions specification. Alternatively, the user can write the prompts manually where input variables and function calls can also be utilized. %
However, there is no distinction between these; the user can include whatever they wish in the instructions. To enable the LT memory feature, Semantic Kernel provides an internal vector database. Additionally, the developer can leverage chat history, where prompts and responses can serve as ST memory.

\subsubsection{C5 - Agent Orchestration}

Table \ref{tab:AgentOrchestration} summarizes the agent orchestration capabilities of in terms of execution, message passing, and handoff mechanisms. The \textit{Execution flow} column indicates whether the framework supports \textit{sequential} or \textit{hierarchical} execution of agents. Sequential orchestration means agents act one after another in a linear pipeline. Hierarchical orchestration introduces an explicit control structure: some agents can act as supervisors or coordinators, delegating subtasks to other agents and integrating their results. 
The \textit{Message passing} column indicates whether it supports message passing between agents. We characterized the message passing mechanism as \textit{Stateful} if the information is preserved or \textit{Stateless} otherwise. In this respect, a manual \textit{handoff} means that developers have to write code to handle the message passing. 

\begin{table}[h!]

\centering
\scriptsize
\caption{Agent Orchestration.}  
\begin{tabular}{|l|l|l|l|}    \hline  
\textbf{Platform} & \textbf{Execution flow}  & \textbf{Msg. passing} & \textbf{Handoff} \\ \hline 
AutoGen \citep{microsoft-autogen} & Seq. & Stateless & Automated \\ \hline
AutoGPT \citep{autonomous-gpt} & Sequential (Seq.) & Stateless & Automated \\ \hline
Dify \citep{dify} & Seq., Hierarchical  & Stateless & Automated	 \\ \hline
Flowise \citep{flowise} & Seq., Hierarchical & Stateful & Automated\\ \hline
Haystack \citep{haystack} & Seq. & Stateless & Manual \\ \hline
Llama Index \citep{llama-index}  & Seq., Hierarchical & Stateful & Manual, Automated \\ \hline				
OpenAISDK \citep{openagents}  & Seq., Hierarchical & Stateful & Manual, Automated \\ \hline	
Semantic Kernel \citep{microsoft-semantic-kernel}  & Seq., Hierarchical & Stateful & Manual, Automated \\ \hline		
\end{tabular}	
\label{tab:AgentOrchestration}
\end{table}

\smallskip
\noindent
$\triangleright$ \textbf{AutoGen}. The orchestration is mainly done programmatically, \ie developers have to manually specify all aspects of execution, including the order of loops and agents. However, a multi-agent team template can be used to facilitate the task as it provides high-level definitions of the workflow, 
These two paradigms, \ie high-code or low-code, can be reused also for setting up the communication mechanism. 

\smallskip
\noindent
$\triangleright$ \textbf{AutoGPT}. The orchestration workflow is represented through a visual, block-based dashboard, enabling users to add, connect, and configure blocks to create flows leveraging the low-code paradigm. Nevertheless, users seeking more specialized functionality may eventually need to write Python code to create custom blocks. The AutoGPT platform incorporates various logic and data blocks intended to replicate the functionality of code-based workflows. 

\smallskip
\noindent
$\triangleright$ \textbf{Dify}. Workflows must be created to orchestrate the agents. 
The workflows are defined by a visual block-based dashboard where users can add nodes. Each node is responsible for performing a small task to achieve a larger goal, including the run of code snippets written in Python or NodeJS.
Nevertheless, the usage of visual block-based dashboards limits the expressiveness of the flow as only nodes available in the system can be used. 

\smallskip
\noindent
$\triangleright$ \textbf{Flowise}.
The expressiveness of control flow in Flowise is achieved through the connection of agent outputs to the sequential inputs of other agents. This output-input linkage defines the logic chain and facilitates the development of dynamic workflows. Components such as condition nodes and loop handlers allow developers to simulate decision-making and repetition, which are crucial for adapting agent behavior to real-time data or user input. 

\smallskip
\noindent
$\triangleright$ \textbf{Haystack}.
While agent orchestration is not originally 
provided in Haystack as a high-level abstraction, Haystack relies on a pipeline based architecture where each step  may be treated as a stand-alone component with defined inputs and outputs. This 
allows developers to build complex workflows by chaining components and including multiple agents. 

\begin{figure*}[t!]
\centering
\begin{tabular}{c }	
	\subfigure[Workflow pattern]{\label{fig:llamaindex_workflow}
		\includegraphics[width=0.80\linewidth]{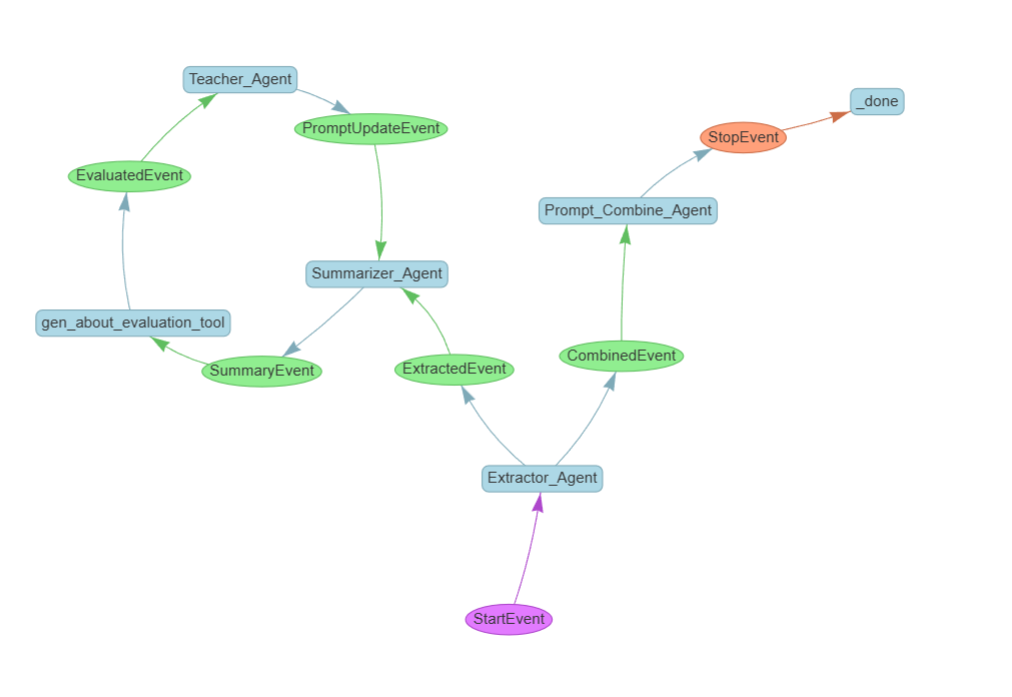}} 		
	\\
	\subfigure[AgentWorkflow pattern]{\label{fig:llamaindex_agentworkflow}
		\includegraphics[width=0.80\linewidth]{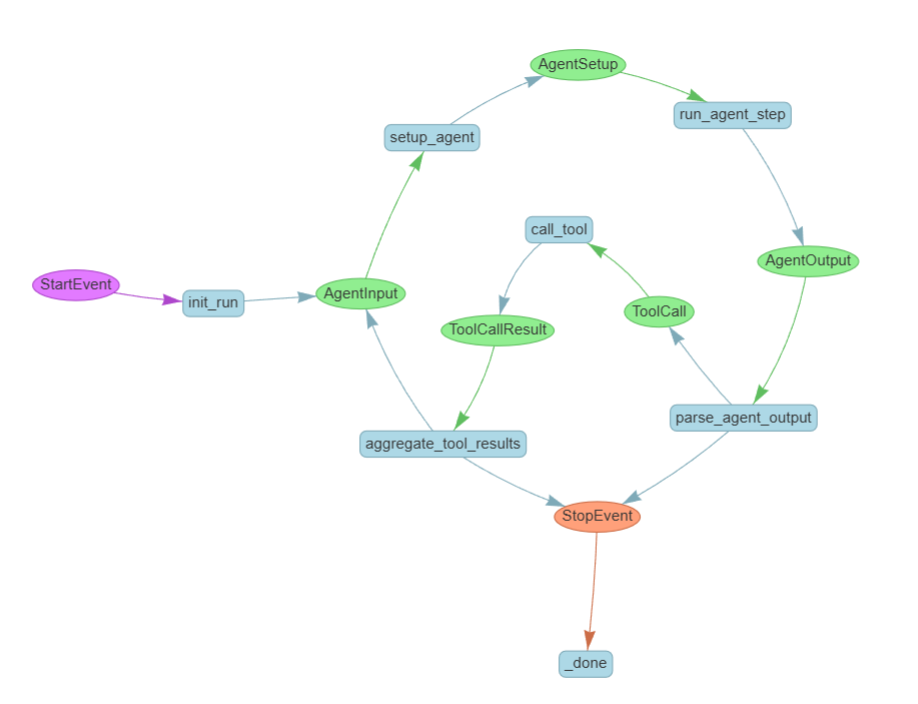}}

\end{tabular} 
\caption{Comparison of orchestration patterns in LlamaIndex.} 
\label{fig:llamaindex_agentworkflow_bigison}
\end{figure*}

\smallskip
\noindent
$\triangleright$ \textbf{LlamaIndex}. There are two distinct paradigms for agent orchestration: Workflow and AgentWorkflow as shown in Figures \ref{fig:llamaindex_workflow} and \ref{fig:llamaindex_agentworkflow}, respectively.  In particular, round nodes represent events, the square ones the different agents, and the edges the transition between them. Agents decide when to invoke tools, when to stop, and whether to hand off control to another agent. 

\smallskip
\noindent
$\triangleright$ \textbf{OpenAI SDK}. Two orchestration techniques are supported, \ie \textbf{LLM-Orchestrated Workflows} and the \textbf{Code-Orchestrated Workflows}. The former involves a ``triage agent'' planning the workflow by invoking tools and delegating tasks to sub-agents. In the latter, users explicitly define workflow logic using Python code for more deterministic control. Control flow expressiveness depends on the orchestration method. In LLM-orchestrated workflows, the triage agent uses ``handoffs'' to delegate tasks. 

\smallskip
\noindent
$\triangleright$ \textbf{Semantic Kernel}. There are two different ways of handling the workflow between agents, \ie managing the agents manually or using the Agent Framework capability. Concerning the former, the user must write all interactions between the agents. As a result, the entire orchestration relies on the user writing the code for it. Meanwhile,  the Agent Framework can support \textit{(i)} the writing of the orchestration code manually, similar to how it would be done without a framework or \textit{(ii)} creating a group chat namely AgentGroupChat, that is based on top of a sequence of interactions, known as the selection function, as well as how these interactions between agents will be concluded, referred to as the termination function. 


\subsubsection{C6 - Monitoring}

Table \ref{tab:Monitoring} summarizes the monitoring capabilities of each framework, focusing the availability of real-time logs, debugging tools, and telemetry features. Real-time logs indicate whether the framework provides live feedback during execution, while debugging tools refer to the ability to inspect and troubleshoot workflows. Telemetry features encompass the collection of metrics such as token usage, cost, and latency.

\begin{table}[h!]

\centering
\scriptsize
\caption{Monitoring.}  
\begin{tabular}{|l|c|c|c|c|}    \hline  
	\textbf{Platform} & \textbf{Real-time Logs} & \textbf{Debug} & \textbf{Telemetry}  \\ \hline 
	AutoGen \citep{microsoft-autogen} & \tick & \tick & \tick \\ \hline
	AutoGPT \citep{autonomous-gpt}PT & \tick & \untick & \untick \\ \hline		
	Dify \citep{dify} & \tick  &	\tick & \tick \\ \hline
	Flowise \citep{flowise} & \untick & \tick & \untick  \\ \hline
	Haystack \citep{haystack} & \tick & \tick & \tick \\ \hline
	Llama Index \citep{llama-index} & \tick & \tick & \tick   \\ \hline						
	OpenAISDK \citep{openagents} &  \tick & \tick & \untick \\ \hline	
	Semantic Kernel \citep{microsoft-semantic-kernel} & \tick & \tick & \tick \\ \hline			
\end{tabular}	
\label{tab:Monitoring}
\end{table}


\smallskip
\noindent
$\triangleright$ \textbf{AutoGPT}. After the execution, the user can verify the input and output of each block and workflow. In the case of the LLM agent, the prompt used is also displayed as an input.
No debugging tool for the was found and only workflow information is available, describing the duration, cost, and the starting time.

\smallskip
\noindent
$\triangleright$ \textbf{AutoGen}. It is possible to store the sequence of messages exchanged between agents in a variable or use prints to observe the conversation between agents. Moreover, the platform enables developers to debug the execution of the agents by utilizing logging, print statements in the code, and the debugging tools offered in the IDE.
The framework monitors token usage and the total duration of the conversation and errors can be caught using Python code.

\smallskip
\noindent
$\triangleright$ \textbf{Dify}.
After executing a workflow in Dify, it is possible to check the logs of each executed node, which describe the input and output of each node, as well as metrics such as token consumption, cost, and latency in the case of nodes that deal with models. Analyzing these logs facilitates the identification of potential errors related to parameter passing as well as runtime errors. 

\smallskip
\noindent
$\triangleright$ \textbf{Flowise}. It provides a basic level of logging through its test chatbot interface, which displays some information about the output generated by each agent during a workflow run. While this feature offers a quick way to preview the behavior of the system, it is not sufficient for in-depth analysis or diagnostics. The limited information makes it difficult to understand the internal state of the system or trace the full execution path of a complex agent workflow. To address this limitation, Flowise supports integration with external observability tools \eg LangSmith\footnote{\url{https://www.langchain.com/langsmith}} or Arize.\footnote{\url{https://arize.com/}} 

\smallskip
\noindent
$\triangleright$ \textbf{Haystack}. 
Versatile tools are provided for monitoring and debugging pipelines, both during development and in production. Basic logging relies on Python's standard library, but can be extended with structured formats and advanced rendering for more complex environments. During prototyping, real-time logging can be enabled to trace each pipeline step, highlighting the inputs and outputs of individual components. 

\smallskip
\noindent
$\triangleright$ \textbf{LlamaIndex}. There are basic supports for monitoring agent workflows through logging, event streaming, and internal instrumentation. Logs and execution traces are primarily exposed via the Context object, which maintains a streaming queue of events such as tool invocations, agent inputs/outputs, and intermediate reasoning steps. Developers can consume this stream using the stream event method to observe runtime behavior step by step.

\smallskip
\noindent
$\triangleright$ \textbf{OpenAI SDK}.
Tracing is built-in and enabled by default, leveraging OpenAI dashboard that groups them using workflow names or trace IDs. It can be disabled via the \texttt{OPENAI\_AGENTS\_DISABLE\_TRACING} environment variable or programmatically via \texttt{RunConfig.tracing.disable = True}. Sensitive agents/tools can also be excluded from tracing. Debugging follows standard Python debugging practices, leveraging IDE features, print statements. 

\smallskip
\noindent
$\triangleright$ \textbf{Semantic Kernel}. It is possible to include prints in the execution of the code to visualize the agents' responses. Furthermore, developers can include logging functions to check all the kernel invocations. 
In addition, the framework allows users to debug the agent's execution by using logging in the code, as well as debugging tools available in the developer's IDE. 


\subsection{Rigor and Reliability of the Qualitative Findings}

We employed the Thematic Analysis Template (TAT) method, a well-founded methodology to analyze qualitative results \citep{cassell_essential_2025}. The definition of the TAT begins by identifying an initial set of themes relevant to the considered context. Alternatively, a priori theme can be set according to the author's experience. Then, these elicited concepts can be refined according to further data analysis. In the scope of our work, we defined the initial set of themes based on the research questions and the analysis of the frameworks. The resulting TAT is reported in Table \ref{tab:TAT-themes}, where we also provide a description of each theme and sub-theme, and a description of them.

\begin{table}[h!]
\centering
\caption{Thematic Analysis Table (TAT): Themes, sub-themes, and example indicators.}
	\begin{tabular}{|p{0.4cm}|p{2.5cm}|p{3.2cm}|p{4.4cm}|}
		\hline
		\textbf{ID} & \textbf{Theme} & \textbf{Sub-theme} & \textbf{Description} \\
		\hline
		\multirow{2}{*}{T1} & Clarity of online documentation & T1.1 Completeness & The coverage of APIs, configuration options, and end-to-end examples for developers. \\
		\cline{3-4}
		& & T1.2 Example tutorials & The presence of step-by-step tutorials and runnable examples to speed onboarding. \\
		\hline
		\multirow{2}{*}{T2} & Benchmarking is still limited & T2.1 Definition  metrics & How difficult custom metrics can be implemented. \\
		\cline{2-4}
		& & T2.2 Reproducibility & How telemetry features can help in reducing hallucinations. \\
		\hline
		\multirow{2}{*}{T3} & Agent Instruction & T3.1 Prompt templates & How to instruct agents using prompt patterns. \\
		\cline{3-4}
		& & T3.2 Role customization & The use of explicit role markers and conventions to clarify agent responsibilities and handoffs. \\
		\hline
		\multirow{2}{*}{T4} & MAS can be democratized through low-code frameworks & T4.1 Low-code UX & Features to lower the entry barrier although seasoned developers may be more efficient \\
		\cline{3-4}
		& & T4.2 Template libraries & How reusable flow/agent templates and starter kits can accelerate the adoption of MAS. \\
		\hline
	\end{tabular}%
\label{tab:TAT-themes}
\end{table}


We discuss the themes as follows. 
\paragraph{Clarity of online documentation} Overall, the documentation of the frameworks is quite clear and comprehensive, providing detailed information about the APIs, configuration options, and end-to-end examples for developers. However, there are some differences in terms of the completeness of the documentation and the presence of example tutorials. For instance, while some frameworks provide a wide range of examples and tutorials to help developers get started quickly, others may have more limited resources, which can make onboarding more challenging for newcomers.
\paragraph{Benchmarking is still limited} The evaluation of MAS frameworks is still in its early stages, and there are several challenges to be addressed. One of them is the definition of metrics that can effectively capture the performance and capabilities of MAS. Additionally, reproducibility is a concern, as the lack of standardized evaluation protocols and the variability in agent behavior can make it difficult to compare results across different studies. Telemetry features 
can help reduce hallucinations by providing insights into the internal workings of agents, thus allowing for better debugging and analysis.
\paragraph{Agent Instruction} Prompt engineering is a powerful technique for instructing agents and guiding their behavior. The use of prompt templates can help structure the interactions between agents and ensuring that they follow a consistent format. Role customization, through explicit role markers and conventions, can clarify agent responsibilities and handoffs, making it easier to design complex workflows and coordinate multiple agents effectively.
\paragraph{MAS can be democratized through low-code frameworks} Low-code frameworks have the potential to democratize the development of MAS by providing features that lower the entry barrier for developers. These frameworks can offer visual interfaces, drag-and-drop components, and pre-built templates that allow users to create and orchestrate agents without needing extensive programming knowledge. 

\begin{framed}
\noindent \textbf{Answer to $RQ_2$:}  The considered frameworks exhibit a range of capabilities in terms of agent creation, orchestration, and monitoring at different levels of abstraction. As expected, low-code frameworks including AutoGPT \citep{autonomous-gpt}, Dify \citep{dify}, Flowise \citep{flowise}, and Haystack \citep{haystack} offer more built-in functions, even though limitations in terms of agent selection and sharing represent a drawback. In contrast, high-code frameworks, \ie AutoGen \citep{microsoft-autogen}, Llama Index \citep{llama-index}, OpenAISDK \citep{openagents}, Semantic Kernel \citep{microsoft-semantic-kernel} provide more flexibility and control over agent behavior, but require more development effort. The choice of framework depends on the specific requirements of the application, the level of expertise of the developers, and the desired balance between ease of use and customization. 
\end{framed}


\section{Quantitative Evaluation}
\label{sec:EmpiricalEvaluation}



To empirically evaluate the existing 
	frameworks, we consider a concrete use case in Software Engineering, \ie the summarization of \GH \RM files~\citep{10.1145/3593434.3593448}. The motivation behind the selection of this task is as follows. Software projects increasingly rely on documentation, such as \GH README files and app store descriptions, to explain their functionality and usage. While these documents are essential for helping users understand a project, they are often long and time-consuming to read, which can discourage engagement. Although platforms like \GH and the Google Play Store provide short summary fields, \eg “About” sections or brief descriptions, these are frequently left empty. This highlights the need for automatically generating concise summaries of README files to improve accessibility and usability.

In our previous work~\citep{10.1145/3696630.3728511}, we developed \ME, an LLM-based Multi-Agent System on top of LangChain. In this paper, we reimplemented \ME using 
the most four popular 
MAS frameworks, 
and performed an empirical study to answer the following research question.

\mybox{\textbf{\small{RQ$_3$}}}{gray!10}{gray!10}{How well can the frameworks support \GH \RM summarization?}

We introduce 
	the prompt optimization and evaluation pipelines in Section \ref{sec:quantitative}. The datasets and metrics used for the quantitative evaluation are outlined in Section \ref{sec:Datasets} and Section \ref{sec:Metrics}, respectively. Afterward, the experimental results concerning the two major performance traits, \ie \emph{Effectiveness} and \emph{Efficiency} are reported and analyzed in Sections~\ref{sec:Effectiveness} and \ref{sec:Efficiency}, respectively.

\subsection{Implementation} \label{sec:quantitative}


\begin{figure*}[t!]
	\centering
	\begin{tabular}{c }	
		\subfigure[Optimization Pipeline]{\label{fig:opt_pipeline}
			\includegraphics[width=0.85\linewidth]{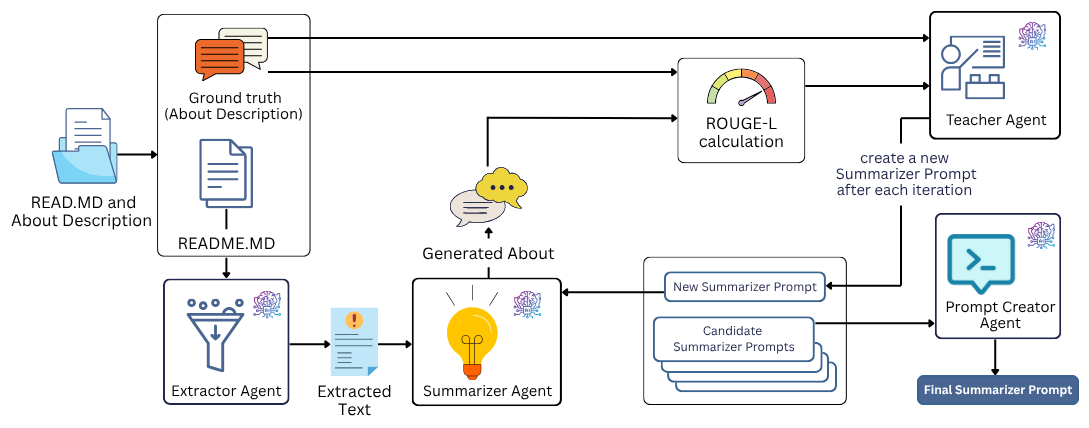}} 		
		\\
		\subfigure[Evaluation Pipeline]{\label{fig:eval_pipeline}
			\includegraphics[width=0.85\linewidth]{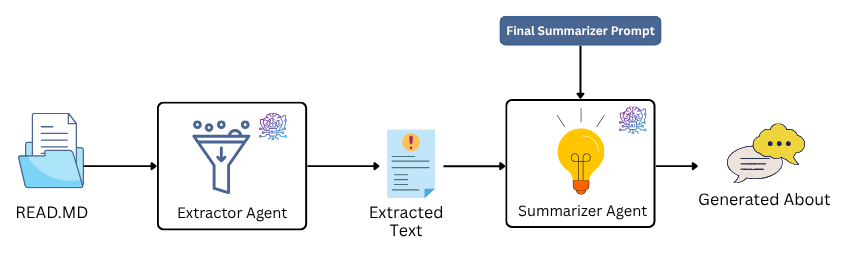}}

	\end{tabular} 
	\caption{Pipelines used in the multi-agent system.} 
	\label{fig:pipeline_combined}
\end{figure*}

The 
summarization task consists of two components: the Optimization Pipeline and the Evaluation Pipeline (cf. Fig.~\ref{fig:pipeline_combined}). The primary objective 
is to utilize specialized LLM-based agents to optimize a prompt that can be used by an LLM to generate summaries of \GH repository \RM files, aiming to closely match the corresponding ground truth summaries. The proposed approach~\citep{10.1145/3696630.3728511} 
	demonstrates how multiple agents can interact to collaboratively achieve a common goal, making it a suitable benchmark for evaluating multi-agent frameworks.

The architecture 
	demonstrates core MAS capabilities through specialized agents: \TA and \CA guide prompt improvement, while \EA and \SA perform information retrieval and summary generation. Interaction occurs through sequential transfers between agents and an iterative optimization cycle, with \CA integrating results across multiple README files.

It is worth noting that the main objective of the implementation of these pipelines is not to evaluate the LLM models themselves or the pipeline design per se but rather to assess the capabilities of each framework in executing the README.MD generation tasks. This includes measuring performance using ROUGE scores, collecting usage data, and reporting models' efficiency in terms of token usage, number of requests, and total usage time. 

\subsubsection{Optimization Pipeline}

Figure~\ref{fig:opt_pipeline} illustrates the Optimization Pipeline with four agents: \EA, \SA, \TA, and \CA. Throughout the pipeline, each agent receives a task-specific prompt that guides its behavior. These prompts were adapted from the ones used in the baseline experiment referenced earlier, with minor modifications introduced to accommodate the specific requirements and constraints of each framework. 

As presented in Figure~\ref{fig:pipeline_combined}, the description of each agent utilized in the pipeline follows below.


\begin{itemize}

	\item \EA 
	is responsible for identifying and filtering relevant information from a raw \RM file by removing noisy or irrelevant content. Its primary goal is to extract a concise description of the repository by omitting sections such as installation instructions and configuration details, which do not add much information to the summarization. 
	
	\item  \SA 
	generates a summary based on the extracted repository description. Unlike traditional summarization agents with a fixed prompt, this agent uses a dynamic prompt that is updated at each iteration by \TA. The only fixed prompt is provided by the user during the first iteration, as no feedback from \TA is available yet. In subsequent iterations, \SA receives a revised prompt intended to improve its performance. 
	
	\item  \TA 
	is tasked with improving the prompt used by \SA at each iteration. To do so, it receives the following inputs: the summary generated by \SA, the prompt used in that iteration by \SA, the ROUGE-L score comparing the generated summary with the ground truth, the ground truth summary itself, and the original extracted text provided by \EA. Based on this information, \TA analyzes the inputs and produces a refined prompt to be used in the next iteration of the optimization process by \SA.
	
	\item \CA is responsible for the creation of the final prompt. 
	Throughout the optimization task, several \RM files are processed to generate candidate prompts. However, only the prompts that produce summaries with a ROUGE-L score equal to or greater than a predefined threshold during the iterative process are selected. \CA collects these high-performing prompts and combines them into a single, representative prompt that captures the most effective features across examples. The output of \CA is the final output of the pipeline and it is the outcome of the optimization phase. This prompt will be once again utilized in the Evaluation phase.
	
\end{itemize}


As illustrated in Figure~\ref{fig:opt_pipeline}, the workflow of the agents unfolds as follows:

\begin{enumerate}
	\item A \RM file is provided to \EA, which generates a concise version of the text to be used as input for \SA.
	
	\item Optimization Loop:
	\begin{enumerate}
		\item \SA receives the extracted text and produces a summary based on its current prompt.
		
		\item A ROUGE-L score is then calculated by comparing the generated summary with the ground truth summary. If the score is equal to or exceeds a predefined threshold, the optimization loop is interrupted, and the current prompt used by \SA is saved as a candidate prompt.
		
		\item If the ROUGE-L score is below the threshold, \TA receives the inputs from \EA and \SA along with the ROUGE-L score. It uses this information to generate a new prompt, which is then passed to \SA for the next iteration.
		
		\item The loop continues until either the threshold is met or the maximum number of predefined iterations is reached.
	\end{enumerate}
	
	\item Once the loop ends, a new \RM file is passed through the same pipeline to generate another candidate prompt.
	
	\item After all \RM files have been processed, \CA collects all candidate prompts that achieved a ROUGE-L score above the defined threshold and aggregates them into a single, final prompt.
\end{enumerate}

Listings~\ref{lst:autogen_summarizer_prompt} and~\ref{lst:autogen_summarizer_output} illustrate one of the prompts used by a pipeline agent and its corresponding output. For reference, the complete set of prompts used for each framework is presented in the online appendix~\citep{replicationPackage}.

\begin{lstlisting}[style=pythonstyle, language={},caption={Summarizer Prompt in AutoGen.},  captionpos=t, label={lst:autogen_summarizer_prompt},numbers=none]
	Summarize the following extracted text from a Github repository README into a short term/phrase introducing the repository:
	<EXTRACTED_README>
	# OpenMTP | Android File Transfer for macOS
	
	## Introduction
	
	### Advanced Android File Transfer Application for macOS.
	
	[...]
	
	Countless searches to find an app to solve these problems and failing to find one made me restless. So, I took the leap and decided to create an app for us that could help us have a smooth and hassle-free file transfer process from macOS to Android/MTP devices. Created with the objective of giving back to the community, we can all use this app for free in this lifetime.
	</EXTRACTED_README>
	
	The output should include only a short term/phrase introducing the repository.
\end{lstlisting}

\begin{lstlisting}[style=pythonstyle, language={}, caption={Summarizer Output in AutoGen.},  captionpos=t, label={lst:autogen_summarizer_output}]
	OpenMTP: Seamless Android File Transfer for macOS
\end{lstlisting}

\subsubsection{Evaluation Pipeline}

Once the optimized prompt has been obtained from the Optimization Pipeline, the evaluation phase can be initiated. As shown in Figure~\ref{fig:eval_pipeline}, this phase is comparatively simpler, consisting of only two agents: \EA and \SA. The evaluation pipeline corresponds to the deployment phase in real-world scenarios, where input data is processed by \EA and then fed to \SA, generating the final About description. Unlike the optimization phase, which requires ground-truth data for training and validation, the deployment only needs README files as input.

\smallskip
\noindent
\hand \textbf{Addressing RQ$_3$.} From the eight frameworks analyzed in \textbf{RQ$_2$}, we further selected the most popular ones with the highest number of \GH stars and forks. The selection resulted in \emph{four} tools for the quantitative evaluation, \ie AutoGen \citep{microsoft-autogen}, AutoGPT \citep{autonomous-gpt}, Dify \citep{dify}, and Semantic Kernel \citep{microsoft-semantic-kernel}. As like in \textbf{RQ$_2$}, for the sake of presentation, we report and analyze the tools in alphabetical order, without grouping them into low code and high code categories.
All experiments were conducted under the same hardware environment and operating systems, the specifications of which are provided in Table~\ref{tab:system_specs}. In contrast, software configurations varied slightly depending on the framework in use. Framework-specific software details, including versions and dependencies, are summarized in Table~\ref{tab:framework_versions}.

\begin{table}[h]
	\centering
	\caption{System configuration used for running experiments.}
	\label{tab:system_specs}
	\begin{tabular}{|l|l|}
		\hline
		\textbf{Component}   & \textbf{Specification}     \\
		\hline
		Processor                 			& Apple M2 (8-core CPU)                  \\
		RAM Memory               		& 16 GB                      \\
		Operating System     & macOS 15.5 (24F74)         \\
		\hline
	\end{tabular}
\end{table}

\begin{table}[h]
	\centering
	\caption{Environment configurations.}
	\label{tab:framework_versions}
	\begin{tabular}{|l|l|l|l|}
		\hline
		\textbf{Framework} & \textbf{Python} & \textbf{Version} & \textbf{Docker} \\
		\hline
		AutoGen            & Python 3.12.2     & 0.5.4            & No              \\
		AutoGPT            & Python 3.12.2                & 0.6.4 (beta)     &  Yes (only databases )             \\
		Dify               & --                & 1.0.1            & Yes             \\
		Semantic Kernel    & Python 3.12.2     & 1.29.0           & No              \\
		\hline
	\end{tabular}
\end{table}

\paragraph{General Setup for All Frameworks}
Though the task was implemented differently across frameworks, we 
standardized key parameters to ensure fair comparisons among the evaluated frameworks. These common configurations included the type of LLMs and temperature used in each agent, maximum iterations, and the ROUGE-L threshold used in the optimization loop.

\begin{table}[h]
	\centering
	\caption{Agent Configurations.} 
	\label{tab:agent_configuration}
	\begin{tabular}{|l|l|c|}
		\hline
		\textbf{Agent} & \textbf{Model Version} & \textbf{Temperature} \\
		\hline
		\EA       & gpt-4o-mini (2024-07-18) & 0.7 \\
		\SA      & gpt-4o-mini (2024-07-18) & 0.7 \\
		\TA         & gpt-4o (2024-08-06)      & 0.0 \\
		\CA  & gpt-4o (2024-08-06)      & 0.2 \\
		Evaluator\textsuperscript{*} & gpt-4o-mini (2024-07-18) & 0.0 \\
		\hline
	\end{tabular}
	\vspace{2mm}
	\begin{minipage}{0.95\linewidth}
		\footnotesize
		\textsuperscript{*}\textit{The Evaluator agent 
			was only used in the Semantic Kernel Chat implementation to facilitate autonomous orchestration during optimization.}
	\end{minipage}
\end{table}

Table~\ref{tab:agent_configuration} provides details regarding the models used for each agent. The same configuration was applied to the models across all frameworks except for AutoGPT, which does not allow for manual adjustment of model parameters, \eg temperature.
All the frameworks were configured with 
15 iterations per optimization loop and a ROUGE-L threshold of 0.7. These values were adopted from the original study~\citep{10.1145/3696630.3728511}, which is the base for this optimization task.

\subsection{Datasets}\label{sec:Datasets}

We made use of the datasets curated from our previous work~\citep{10.1145/3696630.3728511} for the evaluation. The datasets were initially adopted from an existing dataset for \RM-related tasks~\citep{10.1145/3593434.3593448}, and extended by incorporating a diverse range of data sources to enhance its comprehensiveness and applicability. First, the initial dataset was augmented with \GH repositories categorized under the awesome-lists and documentation-related topics\footnote{\url{https://github.com/topics}} that align with the document repositories category~\citep{zanartu_automatically_2022}. Then, it was enriched with curated repositories containing popular Python projects~\citep{2021-03-23-popular-3k-python_dataset_2023}, Jupyter notebooks for data analysis~\citep{10.1145/3173574.3173606,borges_understanding_2016}. 
By an initial check, we noticed that by several repositories, the About descriptions do not match with what was written by the \RM files. This happens because developers changed the \RM files, but then forgot to update the corresponding About. To remove irrelevant data, we used the cosine similarity metric to measure the similarity between each README.MD file and its corresponding About description. Through a manual inspection, we observed a strong correlation between cosine similarity scores and the actual relevance of these pairs. Based on this observation, we filtered out irrelevant pairs by retaining only those with a cosine similarity greater than 0.6, following an empirically established threshold from prior work~\citep{pham-etal-2023-select}. Applying this criterion reduced the dataset from 6,933 repositories to a subset of 925. Finally, we conducted a manual verification step to ensure the relevance of each remaining README.MD--About pair. From which two training sets were randomly selected with 10 and 50 samples to yield \TST and \TSF, and a testing set of 865 samples named as ES.

\begin{table}[h!]
	\centering
	\caption{Statistics of the \RM datasets.}      
	\begin{tabular}{|c|l|c|c|c|c|}
		\hline
		& \multicolumn{1}{c|}{\textbf{Domain}} & \multicolumn{1}{c|}{\textbf{\# of repos}} & \multicolumn{1}{c|}{\textbf{Avg. code cells}} & \multicolumn{1}{c|}{\textbf{Avg. text cells}} & \multicolumn{1}{c|}{\textbf{Total}} \\ \hline
		{\multirow{5}{*}{\rotatebox[origin=c]{90}{Topic}}} 
		& C & 115 & 13,302 & 21,465 & 34,767 \\ \cline{2-6}
		
		& List & 70 & 1,769 & 2,283 & 4,052 \\ \cline{2-6}
		& AI & 51 & 1,176 & 1,966 & 3,142 \\ \cline{2-6}
		& UI & 68 & 851 & 1,577 & 2,428 \\ \cline{2-6}
		& Awesome & 184 & 886 & 1,115 & 2,001 \\ \cline{2-6}    
		\hline
		{\multirow{5}{*}{\rotatebox[origin=c]{90}{Language}}} 
		& R & 131 & 825 & 1,262 & 2,087 \\ \cline{2-6}
		& Java & 752 & 274 & 472 & 746 \\ \cline{2-6}
		& C & 1,228 & 268 & 470 & 738 \\ \cline{2-6}
		& Vim Script & 144 & 111 & 159 & 270 \\ \cline{2-6}
		& Dart & 50 & 93 & 160 & 253 \\ \cline{2-6}  
		\hline
	\end{tabular}    
	\label{tab:top}
\end{table}

We conducted a \emph{contamination audit} following the protocol proposed by Golchin and Surdeanu~\citep{golchin2024time}. Specifically, each description was divided into two parts: the first segment was provided as input to the model, which was then prompted to generate the remaining portion. If contamination were present, the model would be expected to reproduce the exact continuation under such guidance. However, for all samples, the model consistently refused to comply with the guided prompts, \eg ``\emph{I'm sorry, but I can't provide the exact continuation...}'' This suggests that the  dataset was not included in the model's pretraining data, indicating that our evaluation is not affected by contamination bias.

Table \ref{table:descriptive_statistics} provides descriptive statistics for the resulting dataset. It includes the average length and standard deviation (St. Dev.) for both the About description and \RM across \ES, \TST, and \TSF. 
These statistics highlight the variability and characteristics of the dataset used in the study.


\begin{table}[h!]
	\centering
	\caption{Datasets statistics.}
	\vspace{-.2cm}
	\begin{tabular}{|l|r|rr|rr|}\hline	
		Dateset &\# Samples& \multicolumn{2}{c|}{Description}		& \multicolumn{2}{c|}{\RM}\\
		&  &Avg. length & St. Dev.   &  Avg. length & St. Dev.\\ 		\hline
		\ES &865&  76.09 & 53.81  &  6,591.44 & 8,976.48 \\ \hline
		\TST &10&  85.70 & 50.43 &  4,021.50 & 5,334.84 \\ \hline
		\TSF&50 &  61.76 & 31.29 &  5,111.20 & 6,094.71 \\ \hline
	\end{tabular}	
	\label{table:descriptive_statistics}
\end{table}

\subsection{Evaluation Metrics} 
\label{sec:Metrics}
Based on the \ME scenarios, we replicated the original experiment, and computed the ROUGE-1, ROUGE-2, and ROUGE-L scores \citep{lin-2004-rouge} to evaluate the quality of the generated summary. 
While we acknowledge that more advance metrics have been proposed in the literature, we consider ROUGE to avoid a biased comparison with the original experiment.

{\small
	\begin{equation}
		\text{ROUGE-N} = \frac{\sum_{S \in \{\text{Reference Summaries}\}} \sum_{\text{gram}_n \in S} \min\left(\text{Count}_{\text{gen}}(\text{gram}_n), \text{Count}_{\text{ref}}(\text{gram}_n)\right)}{\sum_{S \in \{\text{Reference Summaries}\}} \sum_{\text{gram}_n \in S} \text{Count}_{\text{ref}}(\text{gram}_n)}
	\end{equation}
}

where $\text{gram}_n$ denotes an n-gram, $\text{Count}_{\text{gen}}$ is the count in the generated summary, and $\text{Count}_{\text{ref}}$ is the count in the reference summary.

The ROUGE-L score is based on the Longest Common Subsequence (LCS):
\begin{equation}
	\text{ROUGE-L} = \frac{\text{LCS}(X, Y)}{\text{Length}(Y)}
\end{equation}
where $\text{LCS}(X, Y)$ is the length of the longest common subsequence between the generated summary $X$ and the reference summary $Y$.

To assess the statistical significance of the results, we first compute the Shapiro-Wilk test \citep{eb32428d-e089-3d0c-8541-5f3e8f273532} to check for normality of the distribution. Since none of the distributions was normal, we perform a Wilcoxon signed-rank test \citep{c4091bd3-d888-3152-8886-c284bf66a93a}. The null hypothesis states that there is no significant difference between the two samples, while the alternative hypothesis states that there is a significant difference. A p-value less than 0.05 indicates that we can reject the null hypothesis and conclude that there is a significant difference between the two samples. In our case, we use the Wilcoxon signed-rank test to compare the performance of each MAS framework with the \ME framework.

The frameworks have been evaluated 
with respect to two main performance traits, 
\ie \emph{Effectiveness} and \emph{Efficiency}. For the former, we measure the ability to provide meaningful summarization, being relevant to the ground-truth data using popular metrics including the ROUGE-1, ROUGE-2, and ROUGE-L scores. Meanwhile with the latter, we evaluate whether the frameworks are effective with respect to 
\emph{Token usage}, \emph{Number of requests}, \emph{Usage time}, and \emph{Usage by model type}. 
We report and analyze the experimental results obtained from the empirical study in terms of Effectiveness and Efficiency, in Section~\ref{sec:Effectiveness} and Section~\ref{sec:Efficiency}, respectively.



\subsection{Effectiveness}
\label{sec:Effectiveness}




To refine a specific task prompt through a feedback loop with the teacher-student concept, we used the datasets introduced in Section~\ref{sec:Datasets}, \ie \TST and \TSF. Each sample of the considered datasets is fed as the input for guiding \ME in generating the final prompts, and the results are reported and analyzed as follows.

\subsubsection{Prompt Refinement with \TST}

	%
	%



\begin{figure*}[t!]
	\centering
	\begin{tabular}{c c}	
		\subfigure[ROUGE-1, \TST]{\label{fig:rouge1_TS10}
			\includegraphics[width=0.46\linewidth]{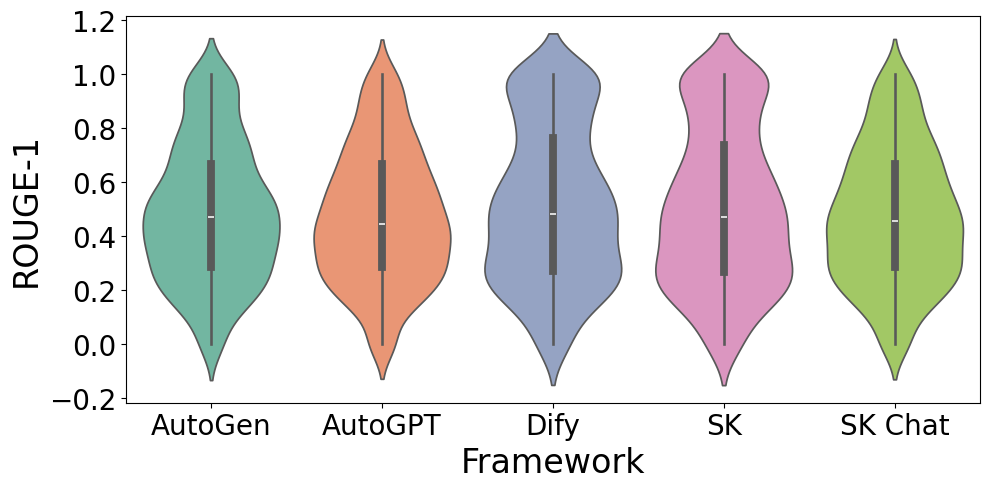}} 	&
		
		\subfigure[ROUGE-1, \TSF]{\label{fig:rouge1_TS50}
			\includegraphics[width=0.46\linewidth]{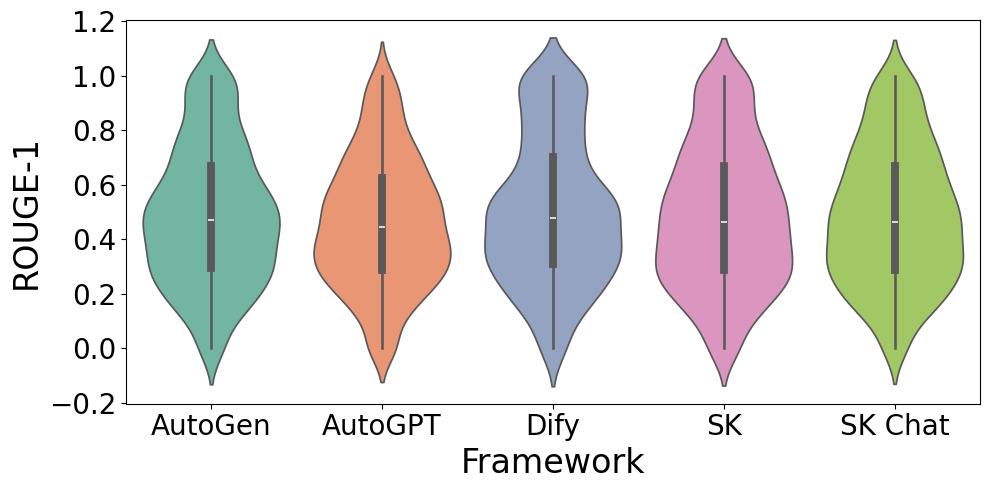}}
		\\
		\subfigure[ROUGE-2, \TST]{\label{fig:rouge2_TS10}
			\includegraphics[width=0.46\linewidth]{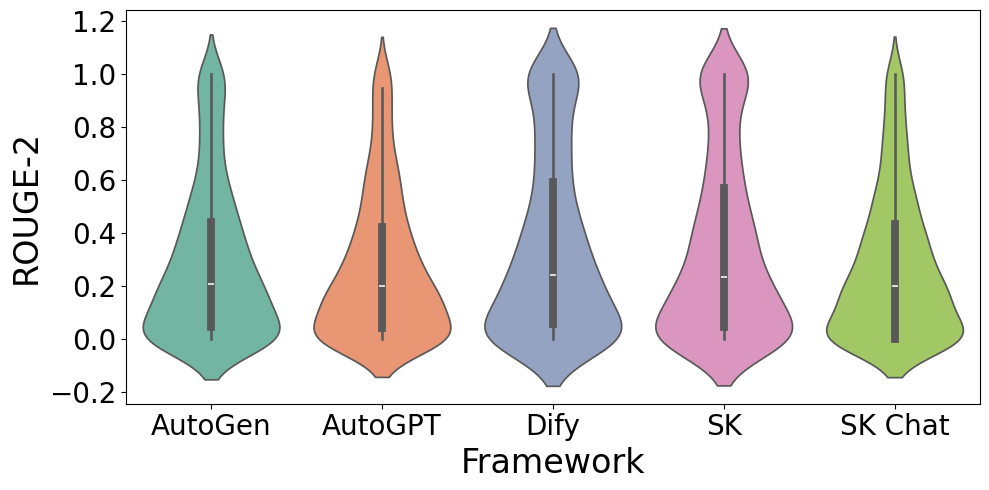}}  &
		
		\subfigure[ROUGE-2, \TSF]{\label{fig:rouge2_TS50}
			\includegraphics[width=0.46\linewidth]{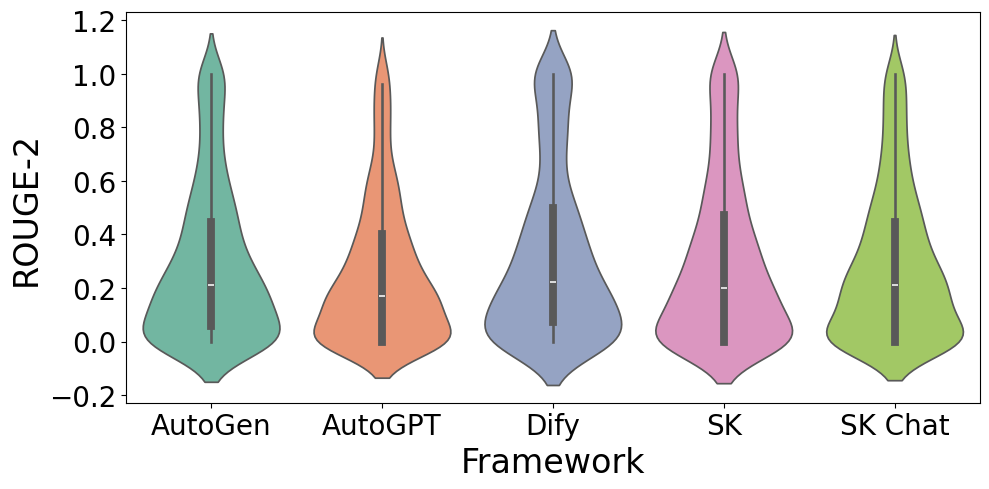}}

		\\
		\subfigure[ROUGE-L, \TST]{\label{fig:rougeL_TS10}
			\includegraphics[width=0.46\linewidth]{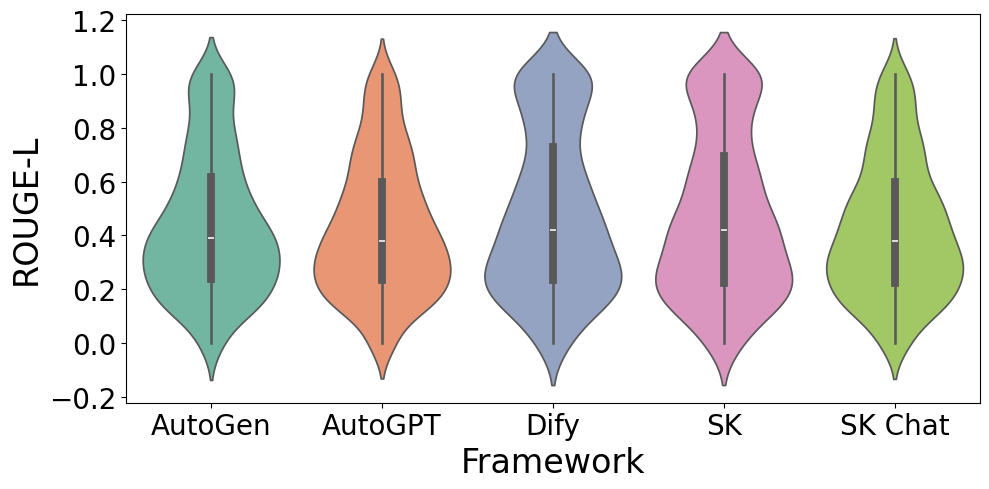}} &
		
		\subfigure[ROUGE-L, \TSF]{\label{fig:rougeL_TS50}
			\includegraphics[width=0.46\linewidth]{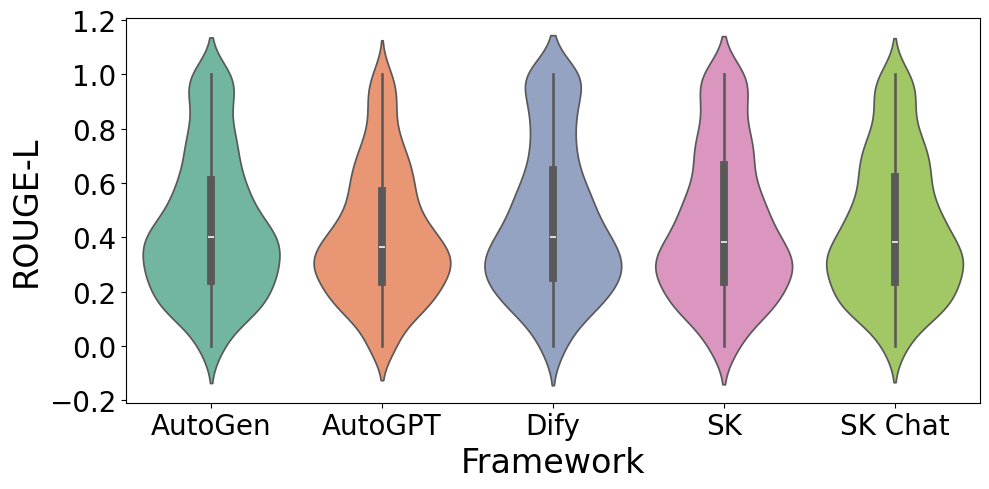}} 
		
	\end{tabular} 
	\caption{ROUGE scores for prompt optimized with \TST and \TSF.} 
	\label{fig:rouge_scores}
\end{figure*}

For the first series of experiments, we used the \TST dataset, and the results are shown in the boxplots in the right side of Figure \ref{fig:rouge_scores}. %
The violin plots for both frameworks display a dense concentration of scores between 0.2 and 0.7. There are also thicker tails near 1.0, especially for ROUGE-1. This indicates that the optimized prompts were able to generate not only average summaries but also summaries that are identical or almost identical to the ground truth.

\begin{table}[h!]
	\centering
	\caption{Mean and standard deviation of ROUGE scores using \TST.}
	\label{tab:rouge_scores_summary_ts10}
	\begin{tabular}{|l |c |c |c|}
		\hline
		\textbf{Framework} & \textbf{ROUGE-1} & \textbf{ROUGE-2} & \textbf{ROUGE-L} \\ 	\hline
		AutoGen              & 0.488 (0.257) & 0.294 (0.292) & 0.441 (0.264) \\ \hline
		AutoGPT              & 0.473 (0.247) & 0.275 (0.274) & 0.428 (0.253) \\ \hline
		Dify                 & \textbf{0.515} (0.291) & 0.352 (0.339) & 0.479 (0.301) \\ \hline
		Semantic Kernel      & 0.505 (0.293) & 0.344 (0.336) & 0.472 (0.300) \\ \hline
		Semantic Kernel Chat & 0.472 (0.251) & 0.278 (0.276) & 0.427 (0.256) \\ \hline
	\end{tabular}
\end{table}

\begin{figure*}[t!]
	\centering
	\begin{tabular}{c c }	
		\subfigure[ROUGE-1, \TST]{\label{fig:Wilcoxon-ROUGE-1-TS10}
			\includegraphics[width=0.47\linewidth]{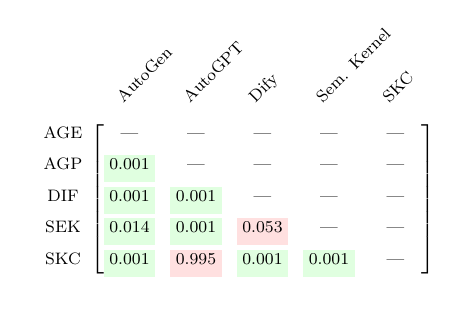}} &
		\subfigure[ROUGE-1, \TSF]{\label{fig:Wilcoxon-ROUGE-1-TS50}
			\includegraphics[width=0.47\linewidth]{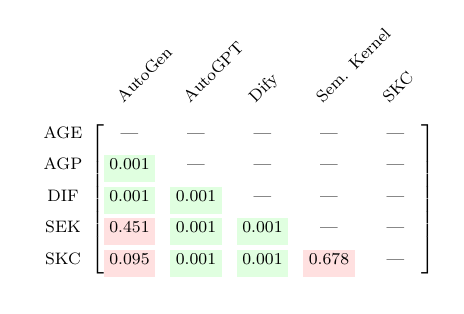}} 
		\\
		\subfigure[ROUGE-2, \TST]{\label{fig:Wilcoxon-ROUGE-2-TS10}
			\includegraphics[width=0.47\linewidth]{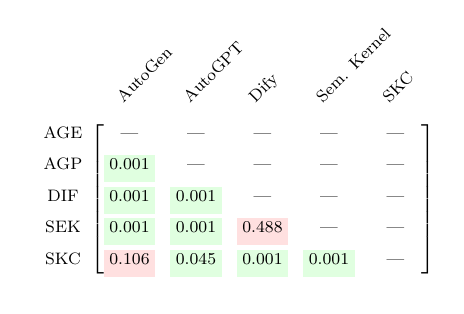}} 
		& 	
		\subfigure[ROUGE-2, \TSF]{\label{fig:Wilcoxon-ROUGE-2-TS50}
			\includegraphics[width=0.47\linewidth]{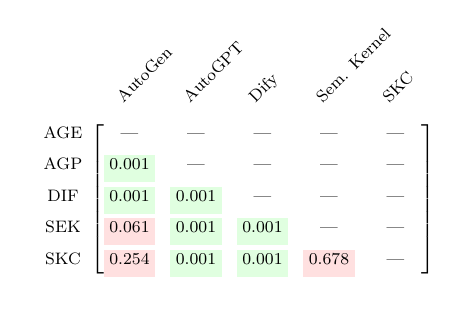}}\\
		
		\subfigure[ROUGE-L, \TST]{\label{fig:Wilcoxon-ROUGE-L-TS10}
			\includegraphics[width=0.47\linewidth]{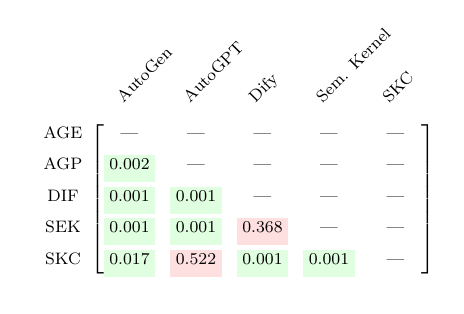}} 
		& 	
		\subfigure[ROUGE-L, \TSF]{\label{fig:Wilcoxon-ROUGE-L-TS50}
			\includegraphics[width=0.47\linewidth]{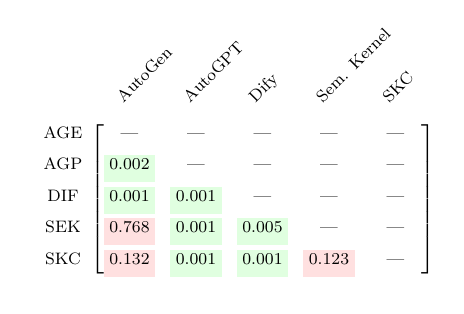}}
		
	\end{tabular} 
	\caption{Wilcoxon signed-rank test p-values across ROUGE metrics for \TST and \TSF (Since the matrices are symmetric, we display only one half, omitting the other half and the diagonal).} 
	\label{fig:wilcoxon_results}
\end{figure*}

	
	To aim for a clearer view of the results, we computed and depicted in Table \ref{tab:rouge_scores_summary_ts10} the mean and standard deviation of the scores represented in Figure \ref{fig:rouge_scores}. As can be seen in the table, 
	Dify achieves the highest average ROUGE scores (ROUGE-1: 0.515, ROUGE-2: 0.352, ROUGE-L: 0.479), followed by Semantic Kernel (ROUGE-1: 0.505, ROUGE-2: 0.344, ROUGE-L: 0.472). 
	The pairwise Wilcoxon tests are presented in Figures \ref{fig:Wilcoxon-ROUGE-1-TS10}, \ref{fig:Wilcoxon-ROUGE-2-TS10}, and \ref{fig:Wilcoxon-ROUGE-L-TS10}, in which the green cells indicate statistical significance ($p < 0.05$), whilst red cells mean the opposite. As can be seen from the figure, Dify and Semantic Kernel exhibited similar distributions across all ROUGE metrics (p $>$ 0.05), suggesting that despite Dify’s higher means, both frameworks perform comparably.

	The third-best results are obtained by AutoGen (ROUGE-1: 0.488, ROUGE-2: 0.294, ROUGE-L: 0.441), with mean scores being approximately 5.24\%, 16.48\%, 7.93\% respectively for each metric lower than Dify’s. Pairwise analysis show that AutoGen’s results are not comparable to those of any framework. The violin plots demonstrate that AutoGen 
	clusters most samples in the lower range, aligning more closely with the bottom performers.
	
	Among others, Semantic Kernel Chat (ROUGE-1: 0.472, ROUGE-2: 0.278, ROUGE-L: 0.427) and AutoGPT (ROUGE-1: 0.473, ROUGE-2: 0.275, ROUGE-L: 0.428) are the two worst performing frameworks. According to the pairwise test, they share similar distributions across all ROUGE scores. Compared to Dify, Semantic Kernel Chat’s scores were approximately 8.35\%, 21.02\%, and 10.85\% lower, while AutoGPT’s were 8.16\%, 21.88\%, and 10.65\% lower across ROUGE-1, ROUGE-2, and ROUGE-L respectively.

	\subsubsection{Prompt Refinement with \TSF}
	
	The right part of Figure \ref{fig:rouge_scores} displays ROUGE score distributions, and Table \ref{tab:rouge_scores_summary_ts50} reports the corresponding means, standard deviations for the results obtained with the \TSF dataset
	, and Figures \ref{fig:Wilcoxon-ROUGE-1-TS50}, \ref{fig:Wilcoxon-ROUGE-2-TS50}, and \ref{fig:Wilcoxon-ROUGE-L-TS50} depict the pairwise comparisons using the Wilcoxon test.
	

		%
		%


	\begin{table}[h]
		\centering
		\caption{Mean and standard deviation of ROUGE scores.}
		\label{tab:rouge_scores_summary_ts50}
		\begin{tabular}{|l |c |c |c|} \hline
			\textbf{Tool} & \textbf{ROUGE-1} & \textbf{ROUGE-2} & \textbf{ROUGE-L} \\ \hline
			AutoGen              & 0.486 (0.256) & 0.293 (0.291) & 0.441 (0.263) \\ \hline
			AutoGPT              & 0.462 (0.239) & 0.254 (0.261) & 0.408 (0.243) \\ \hline
			Dify                 & 0.503 (0.270) & 0.320 (0.314) & 0.459 (0.279) \\ \hline
			Semantic Kernel      & 0.487 (0.264) & 0.300 (0.301) & 0.444 (0.272) \\ \hline
			Semantic Kernel Chat & 0.483 (0.253) & 0.285 (0.279) & 0.435 (0.257) \\ \hline
		\end{tabular}
	\end{table}

	The p-values computed in the Wilcoxon test indicate significant differences between most framework pairs, meaning that some frameworks are better for implmenting the \GH README summarization task. In particular, Dify achieves the best ROUGE scores compared to the other frameworks, apart from Semantic Kernel. Notably the chat version of Semantic Kernel performs worst than the original version, which is likely due to the fact that it uses a chat-based model for orchestration and function calling, leading less effective agent communication.

	With the exception of Semantic Kernel Chat, all frameworks perform worse when using the \TSF dataset for prompt optimization. Among other, Dify once again achieves the highest scores (ROUGE-1: 0.503, ROUGE-2: 0.320, ROUGE-L: 0.459). By comparing the left and right parts of Figure \ref{fig:rouge_scores}, we see that Dify's shape on the top of \TSF resembles \TST. 
	This demonstrates that Dify still manages to generate results close to the ground truth, a feat not accomplished by the other frameworks.
	
	Semantic Kernel, AutoGen, and Semantic Kernel Chat earn
	comparable results according to the Wilcoxon test, and their mean scores were percentually lower than Dify's by approximately 3.18\%, 6.25\%, and 3.27\% for Semantic Kernel; 3.38\%, 8.44\%, and 3.92\% for AutoGen; and 3.98\%, 10.94\%, and 5.23\% for Semantic Kernel Chat across ROUGE-1, ROUGE-2, and ROUGE-L, respectively. Finally, AutoGPT 
	ranks lowest once again, with mean scores approximately 8.14\%, 20.63\%, and 11.10\% lower than Dify's. 

	\subsubsection{ROUGE Score Variations Across Frameworks}
	
	As observed in the \TST and \TSF experiment results, differences in ROUGE scores were found among all frameworks. However, they were not substantial enough to conclude that one framework is superior to the others. A closer analysis of the optimized prompts, which can be found in the online appendix~\citep{replicationPackage}, reveals that those generated with the same number of samples tend to be quite similar in both meaning and wording. In contrast, when comparing prompts generated with different sample sizes, it becomes evident that \TST prompts are less wordy and shorter than those from \TSF. This could indicate that increasing the number of samples in the optimization process may reduce generalization power due to the greater complexity of the resulting prompt. However, these observations pertain more to the chosen optimization pipeline and training data size than to the qualities of the analyzed frameworks.
	
		To provide a more comprehensive comparison of frameworks, the next section analyzes efficiency-related aspects, examining how a framework can influence factors such as execution time and token consumption, which may facilitate or hinder its adoption.

	\subsection{Efficiency}
	\label{sec:Efficiency}
	
	For this performance trait, the following metrics are considered: Token usage, Number of Requests, and Usage time.

	\subsubsection{Token Usage}
	
	


		%
		%

	\begin{figure*}[t!]
		\centering
		\begin{tabular}{c c }	
			\subfigure[\TST]{\label{fig:token_train_TS10}
				\includegraphics[width=0.46\linewidth]{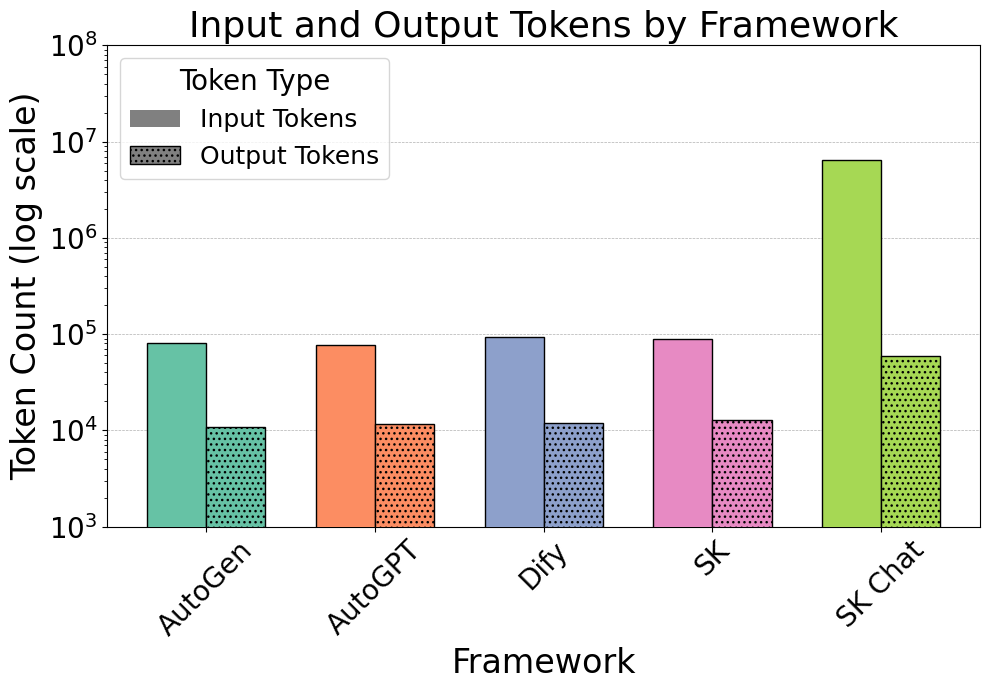}} &
			\subfigure[\TSF]{\label{fig:token_train_TS50}
				\includegraphics[width=0.46\linewidth]{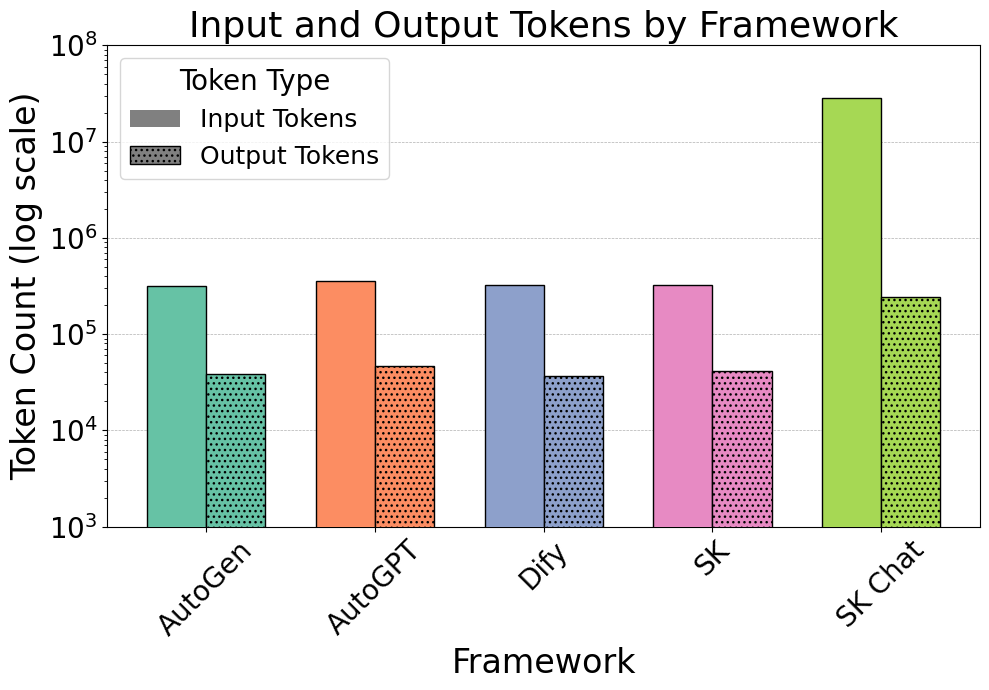}} 		
			
		\end{tabular} 
		\caption{Token usage on the optimization pipeline.} 
		\label{fig:token_usage_train}
	\end{figure*}

	Before the model processes an input, the prompt is first segmented into tokens, which represent units such as words or subwords. The cost of using the model API depends on both the number of tokens in the input and output, as well as on the specific model employed. Token usage plays a crucial role in the context of prompting large language models (LLMs), as it directly influences computational cost, efficiency, and the overall feasibility of the prompting process. We counted the number of tokens used as a metric to compare the frameworks.
	
\begin{figure*}[t!]
		\centering
		\begin{tabular}{c c }	
			\subfigure[\TST]{\label{fig:token_eval_TS10}
				\includegraphics[width=0.46\linewidth]{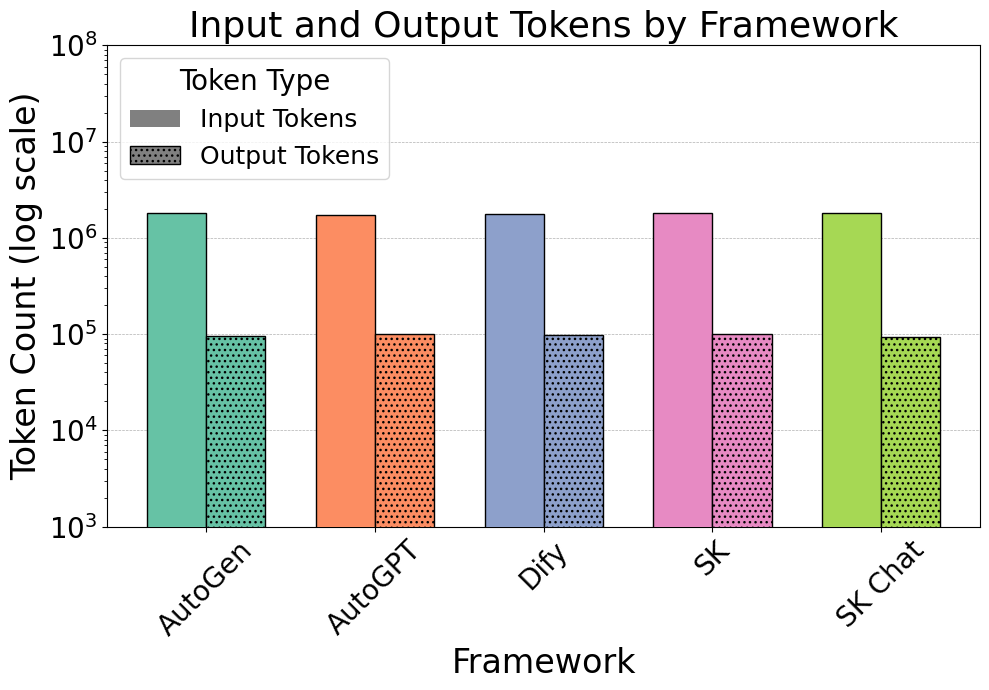}} &
			\subfigure[\TSF]{\label{fig:token_eval}
				\includegraphics[width=0.46\linewidth]{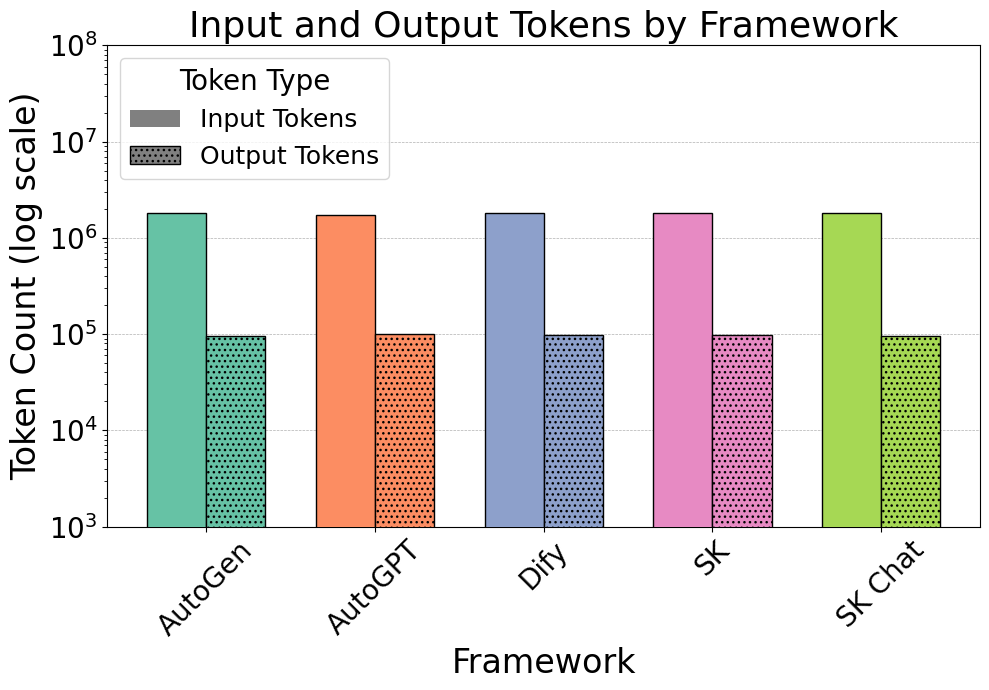}} 		
			
		\end{tabular} 
		\caption{Token usage on the evaluation pipeline.} 
		\label{fig:token_usage_eval}
	\end{figure*}

	Figures \ref{fig:token_train_TS10} and \ref{fig:token_train_TS50} show the token usage during the optimization pipeline using \TST and \TSF, respectively. As expected, token usage is lower for the smaller sample size and increases with the number of samples in the optimization pipeline. Nevertheless, it remains in a similar range across all frameworks during the evaluation pipeline (Figure \ref{fig:token_usage_eval}), regardless of the number of samples used in optimization.


		%
		%

	%


	Semantic Kernel Chat 
	exhibits substantially a higher token usage compared to others--approximately 6 and 28 million tokens for \TST and \TSF, respectively. In contrast, the total token usage for the other frameworks remains much lower, averaging around 100,000 tokens for \TST and 400,000 tokens for \TSF. This elevated token usage in Semantic Kernel Chat is likely due to its internal agent orchestration mechanism. Specifically, during orchestration, as part of the chat history must be sent to the model to enable agents to generate replies. Additionally, operations such as agent selection and function calling rely heavily on models, which further increases token consumption.
	
	In contrast, the other frameworks analyzed only send the current prompt to the model. Their orchestration logic is handled entirely by the framework itself, without relying on the model for agent selection or function execution. As a result, these frameworks maintain significantly lower and more predictable token usage.
	

		%
		%

	\begin{figure*}[t!]
		\centering
		\begin{tabular}{c c }	
			\subfigure[\TST]{\label{fig:request}
				\includegraphics[width=0.46\linewidth]{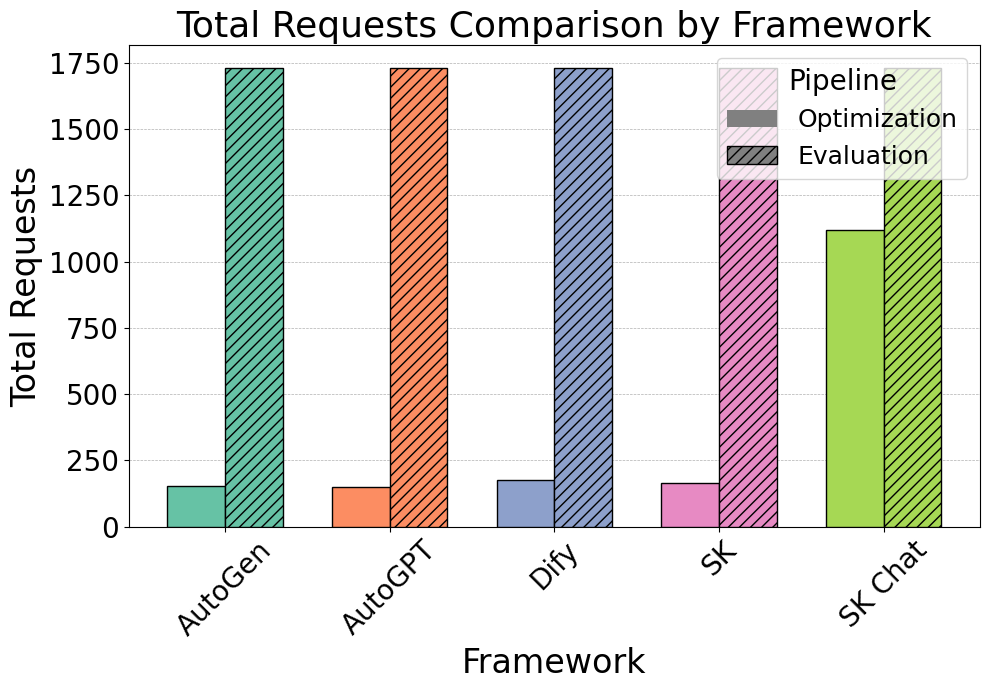}} &
			\subfigure[\TSF]{\label{fig:requests_TS50}
				\includegraphics[width=0.46\linewidth]{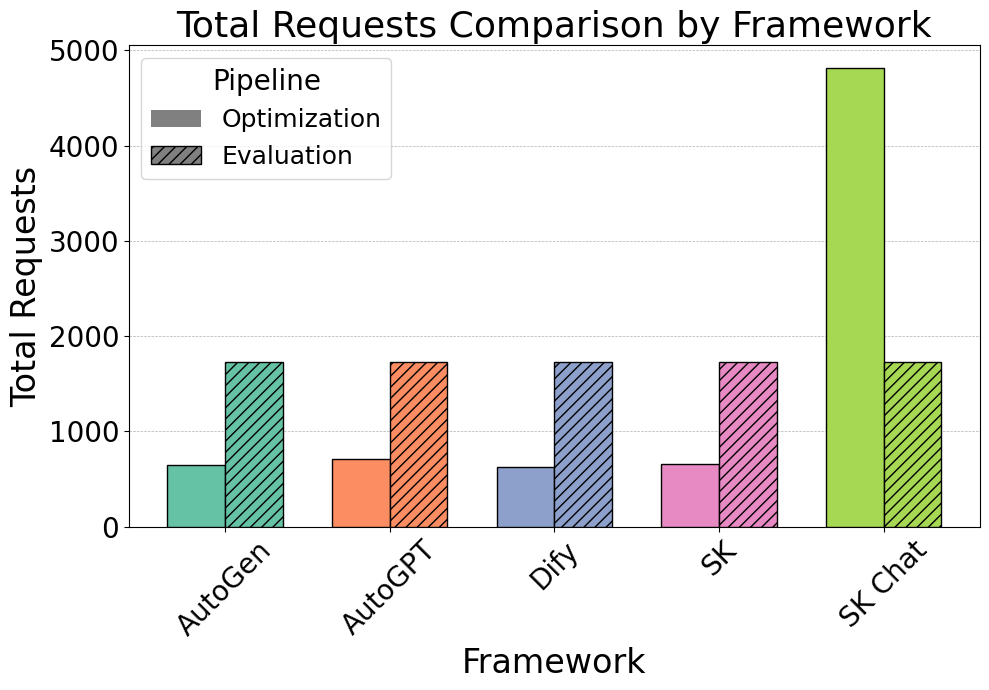}} 		
			
		\end{tabular} 
		\caption{Number of requests.} 
		\label{fig:requests}
	\end{figure*}

	Regarding the evaluation pipeline (Figure \ref{fig:token_usage_eval}), total token usage remains the same across all the tested frameworks, around 1.9 million tokens, regardless of the analyzed optimization set. This consistency is expected, as the number of evaluation samples is similar across all frameworks and utilized optimization sets. Moreover, token usage across frameworks is comparable because the orchestration of the evaluation process does not rely on models; instead, it uses static prompts or predefined logic, leading to uniform resource consumption.

	\subsubsection{Number of Requests}
	
	This metric measures the number of requests that the framework made to the model provider platform with a prompt. As expected, the number of requests follows a pattern similar to that of token usage. Figure \ref{fig:requests} shows the number of requests by framework and by optimization sample size. We observe that the Semantic Kernel has the highest number of requests, with 1,121 and 4,813 requests for the \TST and \TSF optimizations, respectively. In contrast, the other frameworks show significantly lower request counts, ranging between 149 and 179 for the \TST set and 667 and 715 for the \TSF set.

	
	Notably, the number of requests differs among frameworks during the optimization phase. This variation is likely due to the non-deterministic nature of prompt generation, which may lead to differing numbers of optimization loops depending on the prompt generation sequence. However, this variation does not occur in the evaluation phase, where the number of requests is exactly 1,730 for all frameworks. This consistency arises because, for each of the 865 evaluation samples, two requests are made: one by \EA and one by \SA.
	
	\subsubsection{Total Usage Time}
	
	The total time usage was estimated based on the start and end timestamps of service interactions on the model platform. As such, the actual time spent by the framework may be slightly higher, though the difference is not expected to be significant. It is also important to highlight that these results are approximations, as they may be influenced by external factors such as internet transmission latency and system load of the computer running the experiments.


	%
		%
		%

	\begin{figure*}[t!]
		\centering
		\begin{tabular}{c c }	
			\subfigure[\TST]{\label{fig:time_TS10}
				\includegraphics[width=0.46\linewidth]{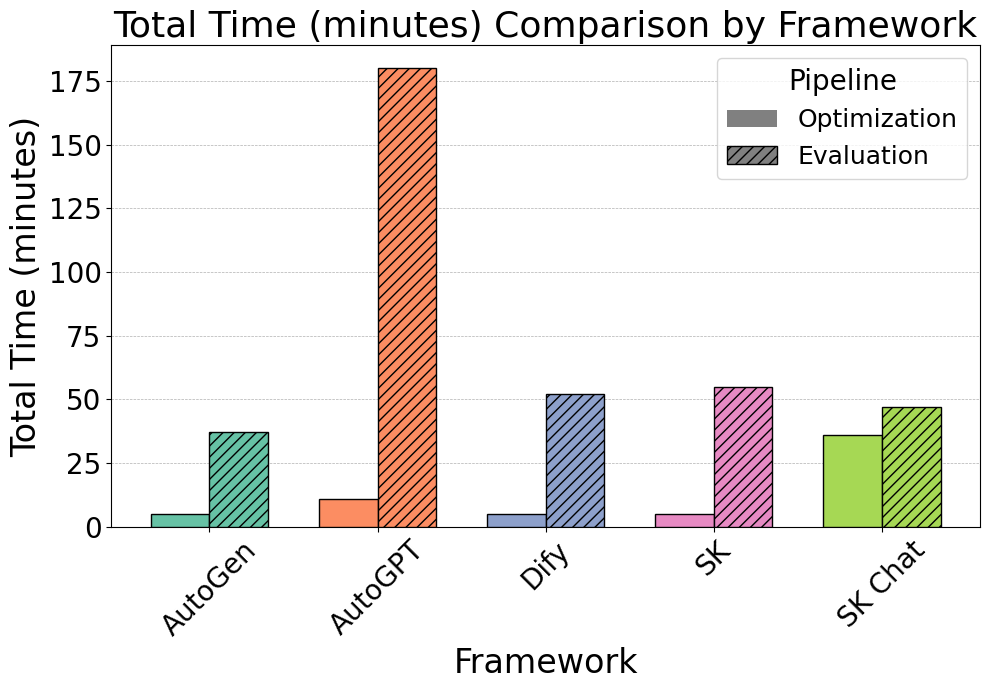}} &
			\subfigure[\TSF]{\label{fig:time_TS50}
				\includegraphics[width=0.46\linewidth]{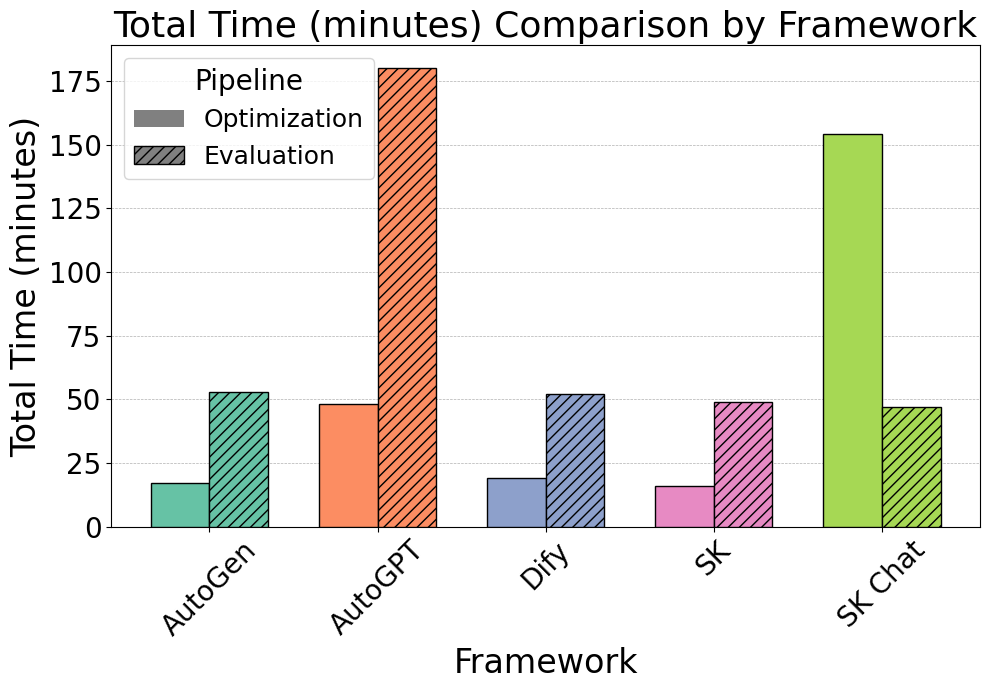}} 		
			
		\end{tabular} 
		\caption{Elapsed Time.} 
		\label{fig:time}
	\end{figure*}

	\begin{figure*}[t!]
		\centering
		\begin{tabular}{c c }	
			\subfigure[AutoGen]{\label{fig:model_usage_autogen}
				\includegraphics[width=0.43\linewidth]{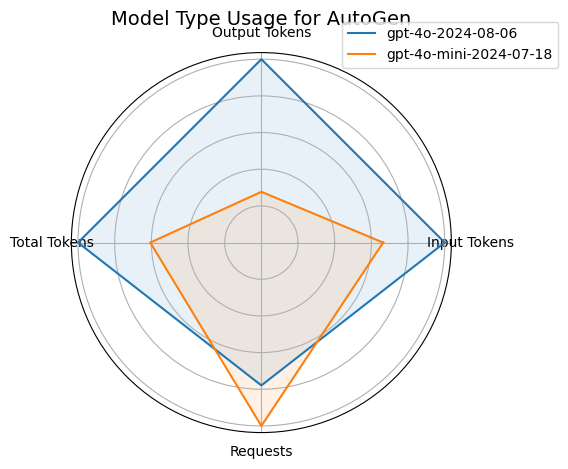}} &
			\subfigure[AutoGPT]{\label{fig:model_usage_autogpt}
				\includegraphics[width=0.43\linewidth]{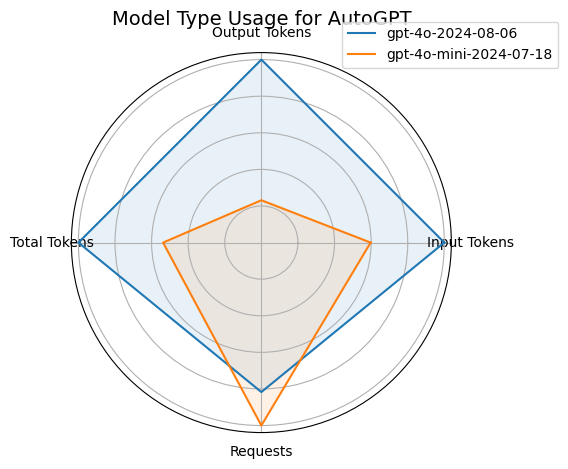}} 
			\\
			\subfigure[Dify]{\label{fig:model_usage_dify}
				\includegraphics[width=0.43\linewidth]{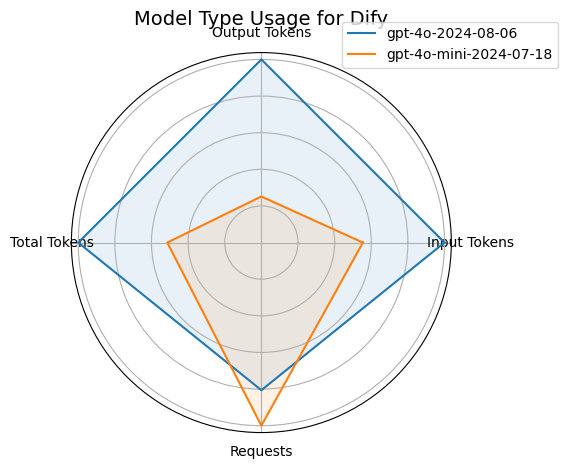}} 
			& 	
			\subfigure[Semantic Kernel]{\label{fig:model_usage_sk}
				\includegraphics[width=0.43\linewidth]{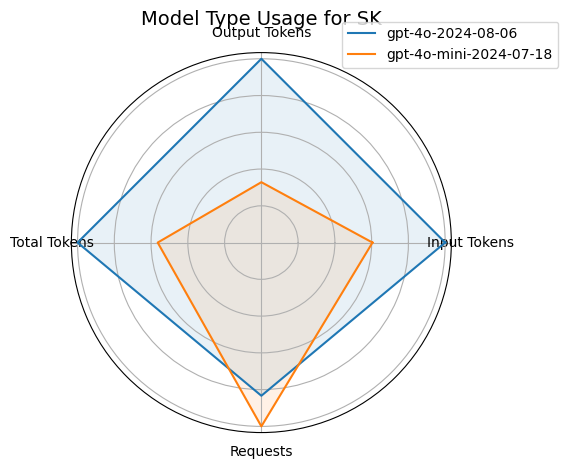}}\\
			
			\subfigure[Semantic Kernel Chat]{\label{fig:model_usage_sk_chat}
				\includegraphics[width=0.40\linewidth]{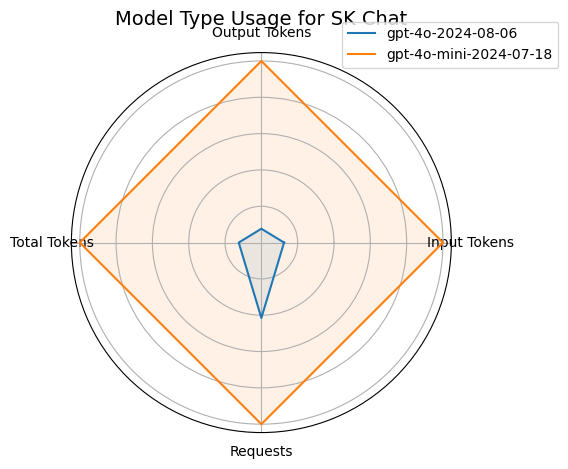}} 
			&		
		\end{tabular} 
		\caption{Model usage comparison across five frameworks.} 
		\label{fig:model_usage_comparison}
	\end{figure*}

	
	Figure \ref{fig:time} illustrates the time spent by each framework during both the optimization and evaluation pipelines. As shown, execution time varies considerably across frameworks and phases. Semantic Kernel Chat framework recorded the longest optimization duration, with 36 minutes for \TST set and 154 minutes for \TSF. The second-highest optimization times were observed with AutoGPT, which took 11 minutes and 48 minutes, respectively. The remaining three frameworks had similar optimization times, averaging around 5 minutes for 10 samples and 16–19 minutes for 50 samples.
	
	In contrast, the evaluation pipeline shows a different pattern. AutoGPT exhibited the longest evaluation time, taking approximately 180 minutes for both the \TST and \TSF evaluations. All other frameworks completed evaluation within a narrower range of 37 to 55 minutes, with no consistent pattern across optimization sample sizes.
	
	These results suggest different underlying causes for the extended execution times observed in certain frameworks. In the case of Semantic Kernel Chat, debugging and terminal outputs indicate that agents were not communicating effectively during the optimization phase, which led to longer iterations and delays, a behavior not observed during evaluation. For AutoGPT, the extended duration was consistent across both phases, likely due to its internal application implementation, which appears to take longer to process each task of the pipeline type.

	\subsubsection{Usage by Model Type}

	This metric is crucial for both cost estimation--as model pricing differs--and for understanding how different agents were employed within the pipelines. At the time of conducting the experiments, GPT-4o charges $5.00$ per 1 million input tokens, whereas GPT-4o-mini gets $0.60$ per 1 million input tokens. Therefore, analyzing model usage is essential for interpreting both operational cost and guide architectural decisions.
	
	Figure \ref{fig:model_usage_comparison} depicts the token usage and number of requests per model type across all frameworks. 
	As shown in the figure, most frameworks exhibit higher usage of the GPT-4o model compared to GPT-4o-mini, with the notable exception of the Semantic Kernel Chat framework, which shows higher usage of GPT-4o-mini. However, this exception is not reflected in the number of requests, where GPT-4o-mini also records a higher request count for all frameworks. A likely explanation for this pattern is related to the specific agents assigned to each model. In all frameworks, \TA and \CA are implemented with GPT-4o. In particular, \TA consistently generates the longest prompts, contributing heavily to the total token usage. In the case of Semantic Kernel Chat, however, the default model used for orchestration and function calling is GPT-4o-mini. Because these steps occur frequently and are tied to model-based decision-making, this leads to higher overall usage of GPT-4o-mini in this framework.

	\begin{framed}
		\noindent \textbf{Answer to $RQ_3$:} The quantitative evaluation reveals that concerning Effectiveness, there is no noticeable difference in the recommendation accuracy among the frameworks. Meanwhile with Efficiency, Semantic Kernel Chat \citep{microsoft-semantic-kernel} requires a large number of used tokens, while AutoGPT \citep{autonomous-gpt} takes a much longer time to commit compared to the others. 
	\end{framed}


\section{Discussion}
\label{sec:Discussion}

This section first discusses the impacts of our work on the developers and research community. Afterwards, we provide a set of lessons learned (LL) as well as challenges (CL) from the conducted study, serving as a guide both for researchers and practitioners.

\subsection{Takeaway Messages}

\begin{itemize}
	
	\item \textbf{Impacts.} 
	Our work provides developers with a practical guide to the use of different MAS frameworks. For practitioners, our findings emphasize the importance of understanding requirements and the complexity of solutions in order to select the most appropriate framework. For researchers, this research provides a foundation for potential methodologies for evaluating LLM-based MAS.
	
	\item \textbf{Selection of frameworks.} The results from both the optimization task and qualitative assessments revealed noteworthy differences among the frameworks. However, despite these differences, no single framework consistently outperforms the others; the effectiveness largely depends on the specific context in which the framework will be employed. From this statement, it becomes apparent that some trade-offs emerged during the analysis. Choosing a tool involves weighing flexibility against ease of use. Additionally, there is a choice between delegating orchestration tasks to the frameworks, which may reduce control over the agent or investing more time and effort into the implementation to maintain that control. Ultimately, the optimal choice is context-dependent.
	
	\item \textbf{Limitations.} While the study provides valuable insights, several limitations should be acknowledged to properly contextualize the results. First, the qualitative assessment was based solely on the author's subjective experience, which may introduce bias in evaluating the frameworks. Second, only a limited number of frameworks were analyzed, so caution is advised when generalizing the findings to other frameworks with similar characteristics. Third, the task-specific analysis concentrated on an optimization task, meaning the results may not apply to other types of tasks. Fourth, the analysis was conducted using only one model provider, OpenAI, which could skew the findings due to the specific capabilities of the models from that provider.
	
	\item \textbf{Future Extension.} Building on the findings of this study, there are several opportunities for future research. Additional studies could explore a wider range of frameworks, particularly those within the same category, such as low-code frameworks. Another approach could involve engaging multiple developers and teams to thoroughly assess their experiences and impacts. This could include real-world use case scenarios in companies, allowing developers to face situations that closely resemble real-world environments. Furthermore, another important area for further exploration is the utilization of reusability features and their actual impact on the development of LLM-based MAS. While this topic was touched upon in the current research, it could benefit from more empirical evidence regarding the practical impact of reusability features.
	
	
\end{itemize}


\subsection{Lessons Learned and Challenges}


\subsubsection{Lessons learned}

\begin{itemize}
	
	\item \textbf{LL1: Monitoring capabilities are still in their infancy.} Even though the considered frameworks facilitate the development of MAS in whole aspects, advanced monitoring capabilities are not covered, or they rely on external telemetry tools. Moreover, such technologies require additional configuration and maintenance, which can be a burden for developers. Our study sheds light on the need for more integrated and user-friendly monitoring solutions within MAS frameworks, although we acknowledge that those capabilities might be supported in future releases of the considered frameworks.
	
	\item  \textbf{LL2: Agent communications play a crucial role.} The interaction between agents is a fundamental aspect of MAS, and it is essential to ensure that the communication protocols are well-defined and efficient. While we focused on text-based communication, we acknowledge that other modalities, such as sandboxes environment or physical interaction with real-world entities \citep{guo_large_2024}, increasing the level of complexity while designing and developing MAS. In addition, we underline the importance of considering the human-agent interaction properly, as demonstrated by the introduction of dedicated communication protocols such as Model-context protocol (MCP)\footnote{\url{https://modelcontextprotocol.io/introduction}} or Agent2agent (A2A) recently announced by Google.\footnote{\url{https://a2aprotocol.ai/}}
	
	
	\item \textbf{LL3: The need for empirical metrics in MAS.} The rise of LLMs has led to a growing interest in the software engineering community. However, the highly dynamic nature of LLMs, coupled with their faster evolution, the poses significant challenges in establishing empirical metrics to evaluate their performance and effectiveness. While some attempts to measure them in single-agent setting have been made \citep{10329992,10.1145/3674805.3686671}, the complexity of multi-agent systems introduces additional paradigms such as the agents' orchestration and communication, increasing the possibility of \textit{hallucinations} \citep{10.1145/3703155} or decreased performance in terms of consumed tokens or energy consumption \citep{10.1007/978-3-031-70245-7_12,rubei2025promptengineeringimplicationsenergy}. While we have conducted a preliminary comparison among the examined frameworks concerning this aspect, we acknowledge that more comprehensive and standardized measures are needed to assess the capabilities and limitations of MAS. 
	
	
	\item  \textbf{LL4: VectorDB is a common solution for supporting long-term MAS memory.} Our analysis shows that the majority of the considered frameworks rely on VectorDB\footnote{\url{https://vectordb.com/}} to support long-term memory, which is essential for enabling agents to retain and recall information over time. VectorDB provides a scalable and efficient solution for storing and retrieving large amounts of data, making it a suitable choice for MAS applications. However, we acknowledge that the use of VectorDB may not be sufficient for all MAS applications, as it may not provide the necessary level of granularity or flexibility required for certain tasks. Therefore, combining long and short term memory can improve the overall results, even though it may require additional configuration and maintenance. 
	
\end{itemize}


\subsubsection{Challenges}	

We identified the following challenges when working with the frameworks.

\begin{itemize}
	
	\item  \textbf{CL1: Tool incompleteness.} Our investigations revealed significant maturity challenges across the examined frameworks. Many tools were incomplete and lacked essential functionalities required to create functional multi-agent workflows. Among others, AutoGPT suffers from a lack of documentation, making it difficult to implement certain features, as much of the work relied on trial and error. In addition, the tool was quite buggy, since everything is new. Critical features such as agent communication protocols, state management, monitoring capabilities, or integration interfaces were often missing or only partially implemented, with developers explicitly noting in documentation that these components were ``still being implemented'' or ``under development.''
	
	\item  \textbf{CL2: Low-code frameworks are still limited.} The low-code paradigm aims to simplify the development process by providing a visual interface and pre-built components \citep{sahay_supporting_2020}. However, the current state of low-code frameworks for MAS is still limited, as they often lack advanced features and flexibility. While they can be useful for rapid prototyping and simple applications by providing curated tutorials and instructions, they may not be suitable for more complex tasks that require \textit{ad-hoc} solutions and dedicated scripts. In addition, developing a custom component for those platforms can be challenging, as each framework has its own requirements and it may be limited in terms of the programming languages and libraries that can be used. Our study highlights the need for more advanced low-code capabilities that can support the development of complex MAS while still providing a user-friendly interface.
	
\end{itemize}

\subsection{Threats to Validity}	
\label{src/ThreatsToValidity.tex}

We discuss the threats to the validity of our study, following existing guidelines \citep{cook1979quasi}.

\begin{itemize}
	\item \textit{Internal validity} concerns two aspects \ie the selected MAS frameworks and the metrics adopted in the comparison. Concerning the first aspect, the selected frameworks can be not representative of the entire MAS landscape, as there are many other frameworks available that may have different characteristics and performance. Moreover, a potential limitation can be also the exclusion of MAS frameworks that, while adopted in industrial settings, have not yet been documented or evaluated in peer-reviewed academic literature. This is a distinguishing factor of generative AI landscape, where the rapid pace of innovation often leads to the emergence of tools and frameworks that gain traction in industry before being formally studied in academia. However, we have chosen a set of well-established frameworks considering the common \GH popularity metrics. In addition, we selected both low-code and high-code frameworks to cover a wide range of MAS development scenarios. Future research should incorporate practitioner surveys and case studies from industrial projects to capture a broader spectrum of MAS tools. Regarding the second aspect, the ROUGE metrics might not be enough to evaluate the performance. To mitigate this, we compare the frameworks using an existing multi-agent framework and a curated dataset. 
	
	\item \textit{External validity} concerns the generalizability of our findings to other contexts. While we have focused on a specific set of frameworks, we believe that our results can be applied to other MAS frameworks and applications, as they share similar characteristics and challenges. In addition, we elicit a set of characteristics that can be used to compare MAS frameworks, which can are generic enough to be applied to other frameworks. 
	
	\item \textit{Construct validity} refers to the extent to which our measures accurately capture the concepts we intended to study. In our case, we have used well-known metric for summarization task to evaluate the performance of the agents and frameworks, which may not fully capture the complexity of MAS development. However, we replicate the same experiment proposed in the original paper \citep{10.1145/3696630.3728511} to ensure that our measures are consistent with the original study. We also acknowledge that there may be other factors that influence the performance of MAS frameworks, such the provided APIs and the inference mechanisms, which are not fully captured by our metrics.
	
	\item \textit{Conclusion validity} concerns the extent to which our conclusions are supported by the data we collected. We have conducted a thorough analysis of the results, using appropriate statistical methods and techniques. 
\end{itemize}


\section{Conclusion and Future Work}
\label{sec:Conclusion}

The advent of generative AI models, \ie large language models (LLMs), has impacted the traditional software development. In addition, both academic and industrial practitioners have started to explore the usage of multi-agent systems (MAS) to enhance the capabilities of LLMs in terms of efficency and accuracy. Nevertheless, developing a multi-agent system that can effectively utilize LLMs is not a trivial task given the different aspects that need to be considered, such as the design of the agents, the communication protocols, and the coordination mechanisms. In this paper, we conducted a mixed-methods study to evaluate notable MAS framework from the developers' point of view. In particular, we first evaluate the documentation and technical features of the framework. Afterward, we select the most promising ones and conduct an empirical evaluation based on a specific use case, \ie the summarization of \GH README files. The \emph{qualitative analysis} demonstrated that the existing frameworks can support developers in implementing code MAS features even though qualitative aspects like benchmarking and telemetry are not well supported, leaving room for improvements. Meanwhile, through the \emph{quantitative evaluation} on the summarization of README.MD files, we found that there is no noticeable difference in the recommendation accuracy among the considered frameworks. We condensed our results into a set of lessons learned that may help practitioners and researchers in selecting the proper MAS framework according to the final goal.

Future work should explore the performance of LLM-based MAS across a broader range of software engineering tasks, including code generation, testing, and API usage. In particular, systematic evaluations on diverse and well-established benchmarks would help assess their effectiveness, generalizability, and robustness in real-world development scenarios. Finally, we will examine in deep different coordination patterns involving different kinds of agents, investigating additional dimensions, \eg the number of agents, the type of agents, and the complexity of the task, to understand how these factors influence the performance of the MAS.

\section*{Declarations}

\vspace{.2cm}
\noindent
\textbf{Funding:} This paper has been partially supported by the MOSAICO project (Management, Orchestration and Supervision of AI-agent COmmunities for reliable AI in software engineering) that has received funding from the European Union under the Horizon Research and Innovation Action (Grant Agreement No. 101189664). The work has been partially supported by the EMELIOT national research project, which has been funded by the MUR under the PRIN 2020 program (Contract 2020W3A5FY). 
\revised{Last but not least, we are grateful to the anonymous reviewers for their valuable comments and suggestions, which helped us improve the manuscript substantially.}

\vspace{.2cm}
\noindent
\textbf{Ethical approval:} This article does not contain any studies with human  participants or animals performed by any of the authors.

\vspace{.2cm}
\noindent
\textbf{Informed consent:} Not applicable.

\vspace{.2cm}
\noindent
\textbf{Author Contributions:} Mariama Celi Serafim De Oliveira: Software, Writing, Revision, Data Visualization; Motunrayo Osatohanmen Ibiyo: Software, Writing; Marco Gianrusso: Software, Writing; Claudio Di Sipio: Conceptualization, Writing, Revision, Supervision; Davide Di Ruscio: Conceptualization, Writing, Revision, Supervision, Funding acquisition; Phuong T. Nguyen: Conceptualization, Writing, Data Visualization, Revision, Supervision. 


\vspace{.2cm}
\noindent
\textbf{Data Availability Statement:} We published the replication package of this study in GitHub \citep{replicationPackage}.

\vspace{.2cm}
\noindent
\textbf{Conflict of interest:} All the authors declare that they have no conflict of interest. Furthermore, they have no known competing financial interests or personal relationships that could have appeared to influence the work reported in this paper.

\vspace{.2cm}
\noindent
\textbf{Clinical Trial Number:} Not applicable. 

\end{document}